\pdfoutput=1

\documentclass[fleqn,usenatbib]{mnras}

\usepackage{newtxtext,newtxmath}
\usepackage[T1]{fontenc}

\DeclareRobustCommand{\VAN}[3]{#2}
\let\VANthebibliography\thebibliography
\def\thebibliography{\DeclareRobustCommand{\VAN}[3]{##3}\VANthebibliography}

\usepackage{graphicx}	% Including figure files
\usepackage{amsmath}	% Advanced maths commands

\title[Testing Chemical Tagging with LAMOST]{Testing Chemical Tagging with LAMOST: Intrinsic Abundance Dispersion of Subgiant Stars in the Galactic Disk}

\author[Yaqian Wu]{Yaqian Wu$^{1}$\thanks{E-mail: wuyaqian@nao.cas.cn}, Maosheng Xiang$^{1, 3}$, \thanks{E-mail: msxiang@nao.cas.cn},
Gang Zhao$^{1,2}$,\thanks{E-mail: gzhao@nao.cas.cn}, Ruizheng Jiang$^{4}$, Zhuohan Li$^{5}$,
 \newauthor Shaolan Bi$^{3,6}$, Meng Zhang$^{1,2}$\\
$^{1}$National Astronomical Observatories, Chinese Academy of Sciences, Beijing 100101, P.\ R.\ China;\\
$^{2}$School of Astronomy and Space Science, University of Chinese Academy of Sciences,
             Beijing 101408, P.\ R.\ China;\\
$^{3}$Institute for Frontiers in Astronomy and Astrophysics, Beijing Normal University,  Beijing 102206, P.\ R.\ China;\\
$^{4}$Purple Mountain Observatory, Chinese Academy of Sciences, No. 10 Yuanhua Road, Nanjing 210023, People’s Republic of China;\\
$^{5}$Shandong Key Laboratory of Space Environment and Exploration Technology, Institute of Space Sciences, \\
School of Space Science and Technology, Shandong University, Weihai 264209, P.\ R.\ China;\\
$^{6}$Department of Astronomy, Beijing Normal University,
             Beijing 100875, P.\ R.\ China;\\
             }

\date{Accepted XXX. Received YYY; in original form ZZZ}

\pubyear{\the\year{}}

\begin{document}
\label{firstpage}
\pagerange{\pageref{firstpage}--\pageref{lastpage}}
\maketitle

\begin{abstract}
The scatter in elemental abundances among stars of similar age and metallicity reflects chemical inhomogeneity in their birth environments, making abundance scatter a key observable for chemical-tagging studies of Galactic formation and evolution.
Using a large sample of subgiant stars with precise ages and elemental abundances derived from LAMOST low-resolution spectra, we investigate the intrinsic chemical abundance scatter of the low-$\alpha$ thin disk near the solar neighborhood ($7 < R < 10$ kpc). We model the abundance ratio [X/Fe], for each of the 19 elements of concern, as a function of age, [Fe/H], and [Mg/Fe], and deduce the intrinsic dispersions with a forward modelling technique. Our results confirm previous findings that the intrinsic scatters are small, typically $\lesssim0.05$~dex, for light elements (C, Al), $\alpha$-elements (O, Mg, Si, Ca, Ti), and iron-peak elements (Mn, Ni). A dedicated analysis of M67 yields similarly small scatter values for these elements, implying limited discriminatory power from light-element abundances alone. 
In contrast, neutron-capture elements exhibit substantially larger scatters, typically $\gtrsim$0.1 dex, which are significantly larger than those of M67 member stars ($\sim$0.07~dex). In particular, our analysis suggests that the abundance variations of individual neutron-capture elements cannot be explained by a single tracer such as [Ba/Fe].
These findings clarify the utility of neutron-capture elements for chemical tagging and highlight the potential of low-resolution spectroscopy in such studies.
\end{abstract}


%%%%%%%%%%%%%%%%%%%%%%%%%%%%%%%%%%%%%%%%%%%%%%%%%%

%%%%%%%%%%%%%%%%% BODY OF PAPER %%%%%%%%%%%%%%%%%%

\section{Introduction}

The chemical abundance patterns of stars preserve a fossil record of the environments in which they formed. Interpreting these abundance distributions—particularly their intrinsic star-to-star scatter—is crucial for understanding the formation and chemical evolution of the Milky Way. The concept of chemical tagging is based on the assumption that stars born from the same molecular cloud share nearly identical chemical compositions across many elements, thereby defining a unique chemical fingerprint that survives even after dynamical dispersal \citep{2002ARA&A..40..487F, 2010ApJ...721..582B, Ting2012}. The feasibility of this method depends critically on the degree and element-dependence of intrinsic abundance scatter within stellar populations.

Over the past decade, large high-resolution spectroscopic surveys such as APOGEE and GALAH have provided powerful tools for studying chemical homogeneity and abundance dispersion across the Galactic disk \citep[e.g.][]{DeSilva2015,Hayden2015, Hogg2016, Majewski2017,Buder2018,Ness2019,Price-Jones2020,Abdurrouf2022}. Using APOGEE data, \citet{Bovy2016} demonstrated that open clusters like M67 and NGC 6819 are chemically homogeneous at the $\lesssim$ 0.02 dex level for light and $\alpha$-elements (e.g., C, N, O, Mg, Fe), implying efficient mixing in their natal clouds. On Galactic scales, \citet{Ness2019} showed that the observed [X/Fe] variations among field red clump stars can be largely explained by stellar age and metallicity, leaving residual intrinsic scatter below 0.03 dex for most $\alpha$- and iron-peak elements. These studies established a clear empirical baseline for the chemical precision attainable in high-resolution spectroscopy.

More recent studies have extended such analyses to include neutron-capture elements, which serve as sensitive tracers of delayed enrichment from asymptotic giant branch (AGB) stars and r-process events \citep{Ting2012,Price2018,Weinberg2019,Ratcliffe2020,Griffith2021}. Using combined APOGEE and GALAH data, \citet{Ratcliffe2023} found that while light and $\alpha$-elements exhibit uniformly low dispersion ($\sim$\,0.02 dex) across the Galactic disk, neutron-capture elements such as Ba, Ce, and Eu display markedly larger intrinsic scatter ($>$\,0.05 dex) and significant spatial gradients. These findings underscore the diagnostic power of heavy elements in tracing localized chemical inhomogeneities and enrichment timescales. More recently, data-driven analyses based on the SDSS-V Milky Way Mapper have further highlighted the rich information encoded in multi-element abundance patterns. Using decoded abundances, \citet{Ness2026b} identified shared disc enrichment patterns among a wide range of chemical species, demonstrating that much of the observed abundance structure can be described by a small number of common enrichment processes. Building on these results, \citet{Ness2026a} reconstructed the corresponding Galactic enrichment pathways and emphasized the power of large spectroscopic surveys for connecting stellar abundance patterns with nucleosynthetic history. While these studies focus primarily on the global structure of abundance correlations, the present work addresses the complementary problem of quantifying the residual intrinsic star-to-star abundance dispersion after removing the dominant dependencies on stellar age, metallicity, and other physical parameters.

Building on this perspective, \citet{Manea2024} demonstrated that incorporating s- and r-process abundances into multi-dimensional chemical space substantially reduces the number of “chemical doppelgängers” \citep{Ness2018}, thereby improving the separability of co-natal stellar populations. Complementarily, \citet{Sinha2024} provided the most stringent observational constraints to date on open-cluster chemical homogeneity. Using APOGEE and SDSS-V Milky Way Mapper data for 26 clusters, they measured abundances of up to 20 elements and found that most $\alpha$- and iron-peak elements remain chemically uniform at the $\lesssim$\,0.02\,dex level, while for neutron-capture elements and weak lines, the upper limits on the intrinsic scatter remain below approximately 0.2\,dex. Their results confirm that open clusters are chemically homogeneous within current precision limits.

Despite these advances, nearly all existing studies rely exclusively on high-resolution spectroscopy (R $\gtrsim$\,20,000), typically covering only $\sim$ 10$^5$\,stars \citep{Buder2018,Ness2019,Feuillet2019,Guiglion2020}. Moreover, the number of well-measured neutron-capture elements remains limited, often constrained by GALAH data \citep{Buder2022,Molero2023}. As a result, our current understanding of how intrinsic chemical scatter depends on element type, stellar age, and Galactic environment remains incomplete. 
Extending such analyses to low-resolution spectra, which now deliver elemental abundances for tens of millions of stars, provides an opportunity to investigate chemical dispersion on a Galactic scale.

Recent methodological progress has demonstrated that LAMOST low-resolution spectra can yield accurate abundances for $\sim$ 20 elements, calibrated against high-resolution benchmarks \citep{Xiang2019, Zhang2024, Zhang2025}. Combined with precise age estimates ($\sim$\,8\%) derived from Gaia–LAMOST synergy for subgiant and red giant stars \citep[e.g.,][]{Xiang2022, Wu2023}, these datasets open a new window to systematically map the intrinsic chemical dispersion of the Milky Way’s disk population.

In this study, we take the first step toward quantifying intrinsic abundance scatter—and, by extension, the feasibility of chemical tagging—in the low-resolution regime. Using a large, homogeneous sample of subgiant stars with precise ages and multi-element abundances from LAMOST DR9, we measure element-specific intrinsic dispersions as functions of age, metallicity, and [$\alpha$/Fe], with particular attention to neutron-capture species. By comparing these results to those of the open cluster M67 and prior high-resolution benchmarks, we aim to assess both the astrophysical robustness and the chemical discriminating power achievable from large-scale, low-resolution spectroscopy.

This paper is organized as follows. In Section\,2, we describe the stellar sample and data sources. Section\,3 details our methodology for estimating intrinsic abundance scatter, including the calibration of measurement uncertainties using repeat observations from LAMOST DR9. 
Section\,4 presents the main results, including the intrinsic abundance dispersions for stellar populations of different age and metallicity, comparisons with open clusters and dispersion associated with [Ba/Fe]. We discuss the implications of our findings in Section\,5, and summarize our conclusions in Section\,6.

\section{Data}

We utilize the subgiant star sample from \citet{Xiang2022}, which provides precise ages and orbital parameters for 247,000 stars derived by combining LAMOST DR7 spectroscopy with Gaia eDR3 astrometry and photometry \citep{Gaia2021}. Subgiant stars offer two key advantages for this study. First, their ages can be determined with high precision owing to the tight correlation between their position in the Hertzsprung–Russell diagram and evolutionary timescale. The inclusion of accurate Gaia photometry and parallaxes further improves age estimates compared to earlier works, yielding a typical uncertainty of only $\sim$8\%. Second, unlike red giants, whose surface abundances are modified by dredge-up processes, subgiants largely preserve their natal chemical composition, making them excellent tracers of Galactic chemical evolution. Moreover, compared to main-sequence turnoff stars—whose inferred ages can vary substantially among stellar models due to systematics in convective-core treatment and diffusion—subgiant ages are more robust and model-insensitive, thus providing a more reliable foundation for studying chemical abundance dispersion.

Stellar parameters and individual elemental abundances of the subgiant sample stars are taken from the LAMOST DR9 stellar abundance catalog derived with  {\sc DD-Payne} \citep{Zhang2025}. The catalog provides effective temperature $T_{\rm eff}$, surface gravity $\log~g$, iron-abundance [Fe/H], and abundance ratio [X/Fe] for 21 elements, namely C, N, O, Na, Mg, Al, Si, Ca, Ti, Cr, Mn, Ni, Y, Sr, Zr, Ba, Nd, Ce, La, Sm, and Eu, based on LAMOST DR9 low-resolution spectra. For spectra with signal-to-noise ratios (S/N) greater than 50, internal precision of stellar label estimates reaches $\sim$30\,K in $T_{\rm eff}$, 0.07\,dex in $\log~g$, $\sim$0.05\,dex in the majority of elemental abundances, and $\sim$0.1-0.2\,dex in abundances of neutron-capture elements. 

In this work, we focus on the low-$\alpha$ disk, which is defined in the [Fe/H] - [$\alpha$/Fe] space using the same criteria as \citet{Xiang2022}, namely,
\begin{equation}
\begin{aligned}
& J_\phi > 1500 \, \mathrm{kpc \cdot km/s}, \quad [\mathrm{Fe}/\mathrm{H}] > -1, \quad \text{and} \\
& [\alpha/\mathrm{Fe}] 
\begin{cases}
> 0.16, & \text{if } [\mathrm{Fe}/\mathrm{H}] > -0.5 \\
< -0.16 \times [\mathrm{Fe}/\mathrm{H}] + 0.08, & \text{if } [\mathrm{Fe}/\mathrm{H}] \leq -0.5
\end{cases}
\end{aligned}
\end{equation}

 Note that [$\alpha$/Fe] represents the global $\alpha$-enhancement parameter inferred by DD-Payne, primarily constrained by the spectral features of the major $\alpha$ elements (e.g. Mg, Si, Ca, and Ti). This quantity is used only for sample selection and trend modelling, while the individual elemental abundances are analysed separately throughout this work.

Figure\,1 shows the distribution of the selected sub-giant sample in this plane, with the red lines demarcating the thin and thick disk populations. The contours indicate the 50th, 70th, 90th, and 99th percentiles of the sample density. As seen from the figure, the majority of stars lie within the thin disk region, confirming the dominance of the low-$\alpha$ population in our sample. This selection yields a final working sample of 66,583 thin disk stars.

Figure\,2 summarizes the basic properties of the final low-$\alpha$ subgiant sample. The age precision of the selected stars is shown in panel (a), where the median fractional age uncertainty is approximately 5\%, slightly better than the median precision reported by \citet{Xiang2022}. The age precision remains nearly constant over most of the age range, with only a mild increase toward the oldest stars.
Panels (b) and (c) show the age and metallicity distributions, respectively. The sample is dominated by intermediate-age stars with metallicities around [Fe/H] $\sim -0.2$ dex, characteristic of the local thin disc population.
Panel (d) shows the age--metallicity distribution. As expected, the majority of stars follow the general trend that older populations are more metal poor. However, a small population of apparently old stars with near-solar metallicity is also present. Similar stars have been reported in previous Gaia--LAMOST and APOGEE studies \citep[e.g.,][]{Xiang2022,Feuillet2019,Wu2023} and are generally interpreted as the combined result of radial migration and age uncertainties. In particular, 468 stars satisfy $10 < {\rm age} < 13$\,Gyr and $-0.1 < {\rm [Fe/H]} < 0.1$, contributing to the oldest age bins in Figures~5 and~8. Nevertheless, these stars account for less than 1\% of the total sample. Moreover, \citet{Wu2023} showed that the oldest thin-disc population has an age of approximately 9.5\,Gyr, implying that stars assigned substantially older ages are largely consistent with the expected measurement uncertainties. Therefore, the presence of this small population is unlikely to affect the main conclusions regarding the intrinsic abundance dispersions derived in this work.

\begin{figure}
\centering
 \includegraphics[width=0.5\textwidth]{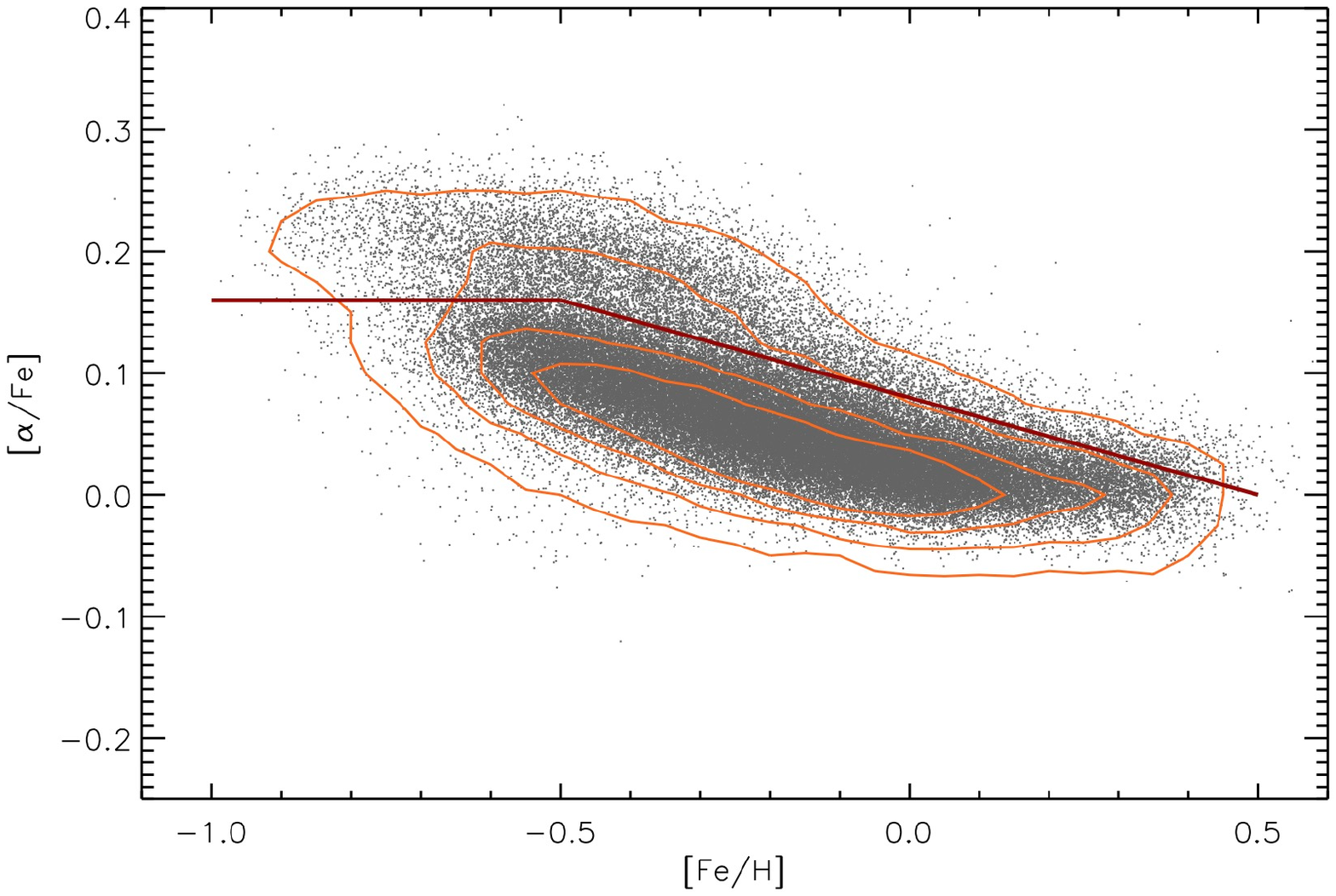}
 \caption{Distributions of the sub-giant stars sample stars in the  [$\alpha$/Fe]–[Fe/${\rm H}$] plane. The red lines delineate the demarcation of the chemical thin and thick disc sequences. The contours indicates 50,70,90, and 99 per cent of the sample stars.}
 \label{fig:fig1}
\end{figure}

\begin{figure*}
    \centering
  \includegraphics[width=0.8\textwidth]{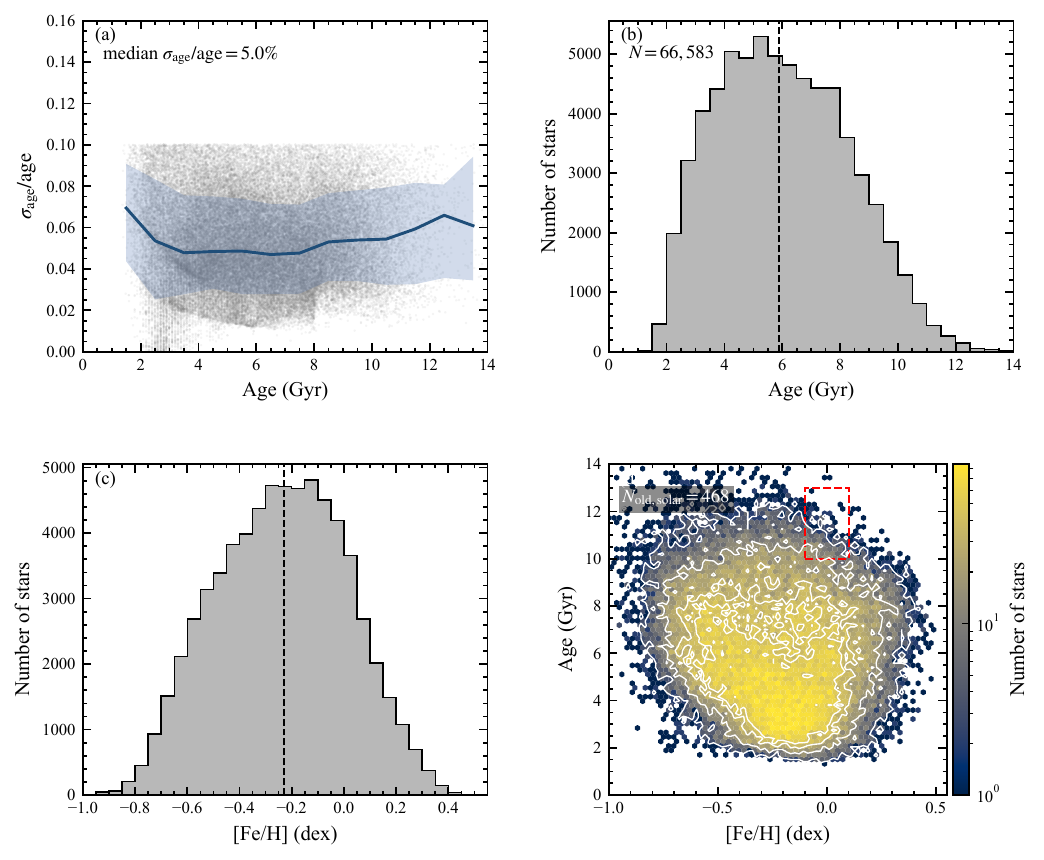}
 \caption{Properties of the final low-$\alpha$ thin-disc subgiant sample. Panel (a) shows the fractional age uncertainty as a function of age. Panels (b) and (c) present the age and metallicity distributions, respectively. Panel (d) displays the age–metallicity distribution, with contours enclosing 50, 70, 90, and 99 per cent of the stars. The red dashed box highlights stars with $10 < {\rm age} < 13$ Gyr and $-0.1 < {\rm [Fe/H]} < 0.1$, containing 468 stars.}
 \label{fig:fig2}
 \end{figure*}

\section{Method}

\subsection{Re-scaling the Abundance Measurement Errors}
Accurate estimation of the measurement errors is crucial for a robust determination of intrinsic abundance dispersion, as the decomposition of total dispersion into intrinsic and observational components is highly sensitive to the assumed error model. To assess the accuracy of the elemental abundance errors reported in the catalog of \citet{Zhang2025}, we examine the abundance difference among repeat observations. About 70\%  of the subgiant stars are observed multiple times (at least 3 times).  Assuming the abundances of a star keep invariant with time, any dispersion among repeat observations reflects the measurement errors. To mitigate biases introduced by differing data quality, we restricted our analysis to pairs of repeat spectra with comparable signal-to-noise ratios, requiring 0.8 $< ({\rm S/N})_1/({\rm S/N})_2 <$ 1.2.

For each element, we calculate the standard deviation of abundance differences between the paired spectra at different spectral S/Ns. This empirical dispersion serves as a proxy for the true measurement errors. Figure\,3 presents the comparison of this empirical abundance dispersion from repeat observations and the measurement errors reported in \citet{Zhang2025} at representative S/N values. While both the empirical and reported errors exhibit the expected dependence on S/N, systematic discrepancies, to a 20\% level, are evident in the abundances of many elements. In particular, the reported errors appear to be overestimated for elements C, O, Mg, Al, Si, Ni, Sr, Y, Nd, Sm, and Eu, whereas they are underestimated for elements including Ti, Ba, and Zr. These differences highlight the need for refined element-specific error models and suggest that incorporating empirical constraints from repeat observations can help improve the precision and reliability of abundance measurements in future data releases.

To reconcile these differences, we derive element-specific correction factors, defined as the median value of ratios of the reported uncertainty to the empirical scatter at various S/N. These correction factors are then applied to recalibrate the abundance errors in the LAMOST abundance catalog. The value of correction factor for each element is marked in the upper-right corner of Figure\,3.  

To further assess the reliability of the DD-Payne abundances for the specific subgiant sample analysed in this work, we cross-matched the sample with GALAH DR4 and APOGEE DR19. Comparisons of stellar parameters and elemental abundances, together with tests for possible temperature- and metallicity-dependent systematics, are presented in Appendices A and B. Overall, the DD-Payne measurements show good agreement with both high-resolution surveys, supporting their use in the intrinsic abundance-dispersion analysis.

\begin{figure*}
 \includegraphics[width=\textwidth]{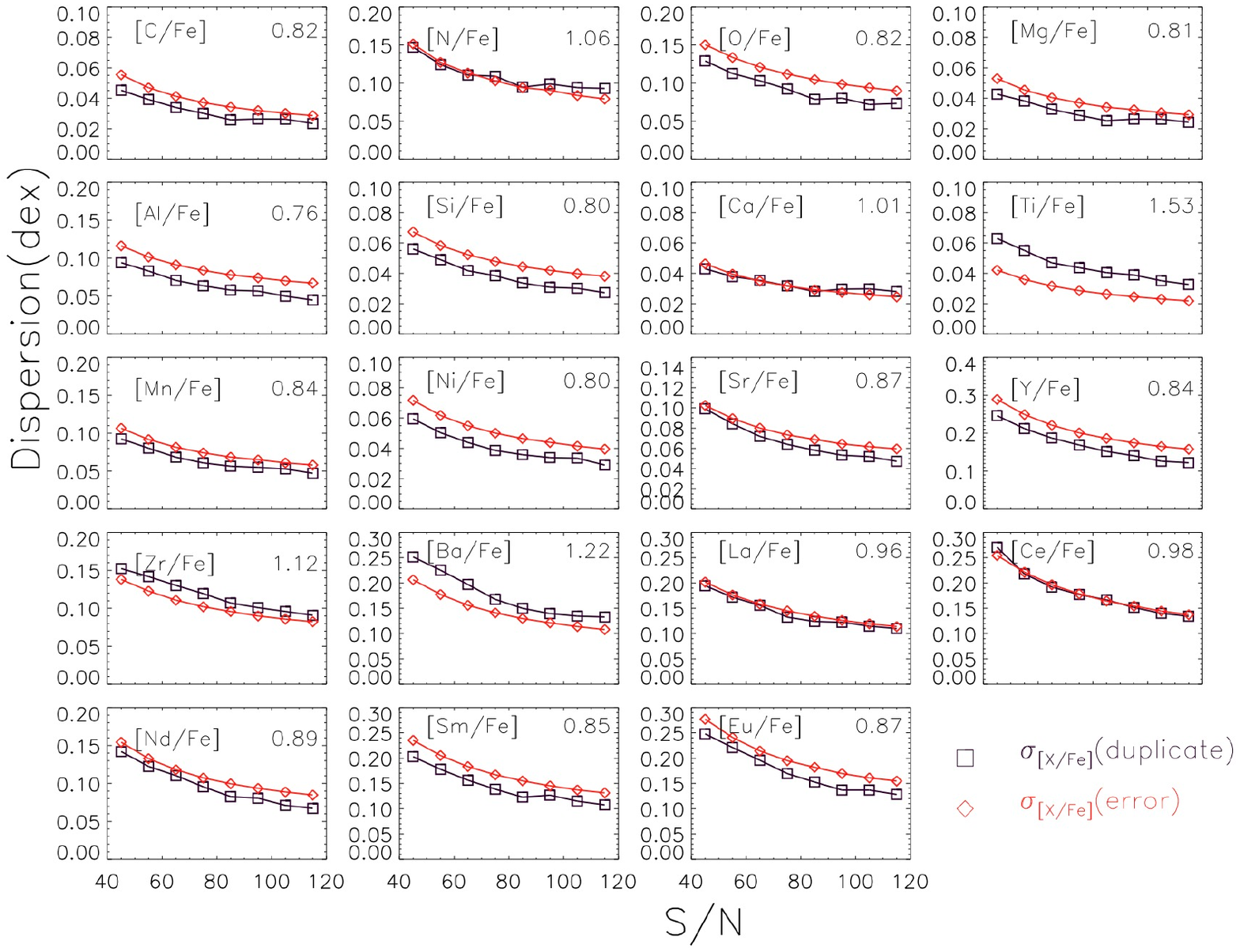}
 \caption{Elemental abundance dispersion as a function of signal-to-noise ratio (S/N) for stars with repeat observations in \citet{Zhang2025}. Black squares represent the standard deviations of abundance measurements derived from paired spectra with similar S/N values (S/N ratio between 0.8 and 1.2). Red diamonds indicate the typical abundance uncertainties reported by \citet{Zhang2025} at the corresponding S/N. The number shown in the upper-right corner of each panel is the scaling factor,
$f=\sigma_{\rm duplicate}/\sigma_{\rm error}$,
where $\sigma_{\rm duplicate}$ is the scatter measured from repeat observations and $\sigma_{\rm error}$ is the corresponding uncertainty reported by \citet{Zhang2025}.}
 %\label{fig:fig3}
\end{figure*}

\subsection{MCMC Method for Determining Intrinsic Dispersion}

The intrinsic dispersion $\sigma_{\rm int}$ of the stellar abundance distribution reflects the extent of chemical inhomogeneity among stars and is a consequence of both chemical enrichment history of the interstellar medium and stellar mixing \citep{Argast2000,2002ARA&A..40..487F}. It is suggested that for the majority of elements $X$, the stellar abundances [$X$/Fe] exhibit a tight astrophysical correlation with -- thus can be well reproduced by -- the combination of a small set of parameters, namely, stellar age $\tau$, metallicity [Fe/H], and [Mg/Fe] \citep{Ness2019, Ratcliffe2023}. Inspired by these studies, we model [$X$/Fe] as a function of a small set of physically motivated quantities, namely stellar age ($\tau$), metallicity ([Fe/H]), [$\alpha$/Fe] enrichment ([Mg/Fe]), s-process enrichment ([Ba/Fe]), and guiding-centre radius ($R_g$). The intrinsic dispersion is inferred from the residual abundance distribution after accounting for the dependence on these conditioning variables.
 
Note that, compared to \citet{Ness2019}, we additionally include [Ba/Fe] as a physically motivated abundance label because Ba is a representative tracer of s-process enrichment. We also include the $R_g$, as an additional parameter to account for potential radial variations in the Galactic abundance distribution.

The impact of incorporating [Mg/Fe] into the model is discussed in detail in Sections 4.1 and 4.2. Furthermore, the role of [Ba/Fe], representing $s$-process enrichment and delayed nucleosynthesis, is examined separately in Section 4.3, where we assess its influence on the inferred abundance trends and intrinsic dispersions.

Our model is expressed as 
\begin{equation}
\overline{\rm{[X/Fe]}}=\rm{[X/Fe]}_{R_{g_{0}}}+\nabla_{\rm{R_{g}}\vert[X/Fe]}(R_{g}-R_{g_{0}})
\end{equation}
where $\nabla_{\rm{R_{g}}\vert[X/Fe]}$ represents the radial abundance gradient,  and the first term on the right is expressed as

\begin{equation}
\begin{aligned}
    \rm{[X/Fe]_{R_{g_{0}}}}=\rm{[X/Fe]_{(R_{g_{0}},\tau_{0},[Fe/H]_{0},[Mg/Fe]_{0},[Ba/Fe]_{0})}}+\\
    \nabla_{\tau\vert\rm{[X/Fe]}}(\tau-\tau_{0})+\nabla_{\rm{[Fe/H]}\vert[X/Fe]}(\rm{[Fe/H]-[Fe/H]_{0})}+\\
    \nabla_{\rm{[Mg/Fe]}\vert[X/Fe]}(\rm{[Mg/Fe]-[Mg/Fe]_{0})}.
\end{aligned}
\end{equation}

For analyses that include [Ba/Fe], the model is expressed as
\begin{equation}
\begin{aligned}
    \rm{[X/Fe]_{R_{g_{0}}}}=\rm{[X/Fe]_{(R_{g_{0}},\tau_{0},[Fe/H]_{0},[Mg/Fe]_{0},[Ba/Fe]_{0})}}+\\
    \nabla_{\tau\vert\rm{[X/Fe]}}(\tau-\tau_{0})+\nabla_{\rm{[Fe/H]}\vert[X/Fe]}(\rm{[Fe/H]-[Fe/H]_{0})}+\\
    \nabla_{\rm{[Mg/Fe]}\vert[X/Fe]}(\rm{[Mg/Fe]-[Mg/Fe]_{0})}+\\
     \nabla_{\rm{[Ba/Fe]}\vert[X/Fe]}(\rm{[Ba/Fe]-[Ba/Fe]_{0})}.
\end{aligned}
\end{equation}

The quantities with a subscript 0, namely $\tau_{0}$, $\rm{[Fe/H]_{0}}$, $\rm{[Mg/Fe]_{0}}$, $\rm{[Ba/Fe]_{0}}$, and $R_{g_{0}}$, represent the median values of the stellar sample.

Finally, for each star i, the observed abundance $\rm{[X/Fe]_{i}}$ is modeled as a realization from a Gaussian distribution centered at the median value  $\overline{\rm{[X/Fe]}}$ and with a dispersion given by the quadrature sum of intrinsic dispersion $\sigma_{\rm int}$ and measurement uncertainties $\sigma_{\rm err}$, $G(\overline{\rm{[X/Fe]}},\sqrt{\sigma_{\rm int}^2+\sigma_{\rm err}^2})$.

Given the measurements $\mathbf{O}:=\{R_{g}, \tau, [Fe/H],[Mg/Fe], [Ba/Fe], [X/Fe]\}$ of the subgiant sample stars, the model parameters $\mathbf{\theta}:= \{\sigma_{\rm int}, \nabla_{\rm{R_{g}}}, \nabla_{\tau}, \nabla_{\rm{[Fe/H]}}, \nabla_{\rm{[Mg/Fe]}}, \nabla_{\rm{[Ba/Fe]}}, Const \} $
can be estimated with a Bayesian scheme
\begin{equation}
P(\mathbf{\theta} | \mathbf{O}) \propto P(\mathbf{O} | \mathbf{\theta})P(\mathbf{\theta}).
\end{equation}
We use a Markov Chain Monte Carlo (MCMC) sampling technique \citep{2013PASP..125..306F} to evaluate the posterior distribution of $\sigma_{\rm int}$ and other parameters, assuming a uniform prior distribution $P(\mathbf{\theta})$.

Following previous studies of stellar abundance dispersions and chemical homogeneity, we assume that both the intrinsic abundance distribution and the observational uncertainties can be approximated by Gaussian functions. Under this assumption, the observed abundance distribution is described by the convolution of the intrinsic scatter and the measurement errors, resulting in a Gaussian distribution with variance $\sigma_{\rm int}^2+\sigma_{\rm obs}^2$. In the absence of intrinsic abundance scatter ($\sigma_{\rm int}=0$), the model naturally reduces to the observational-error distribution alone.
We emphasize that the purpose of the adopted analytical model is not to reproduce the detailed chemical evolution of the Galactic disk, but rather to remove the dominant abundance trends so that the residual intrinsic dispersion can be robustly quantified.

Figure~4 presents an example of the MCMC modelling of the [C/Fe] distribution for stars with $-0.2<{\rm [Fe/H]}<0.2$ and $5<\tau<6$~Gyr. This Figure shows that all the model parameters are well constrained. The covariances among the model parameters are generally weak, suggesting that the adopted labels trace distinct aspects of Galactic chemical evolution. The modelling yields an intrinsic dispersion of 0.03 dex for [C/Fe].

We note that the inclusion of [Mg/Fe] and [Ba/Fe] in the model is motivated by their astrophysical significance rather than by any known measurement covariance. Previous investigations of the DD-Payne abundance labels \citep{Xiang2019,Zhang2025} have shown that the elemental abundances recovered from LAMOST low-resolution spectra are not dominated by strong cross-talk between different labels. Although individual spectral features may be blended, the abundance determination is constrained by information distributed over a large wavelength range, which substantially reduces the impact of local line overlap. Consequently, the correlations involving [Mg/Fe] and [Ba/Fe] are interpreted primarily as signatures of Galactic chemical evolution rather than artifacts of the abundance measurement procedure.

Figure\,5 illustrates the size of the total dispersion, intrinsic dispersion, and measurement error of all the 19 elemental abundances for our sample stars with $-0.2<{\rm [Fe/H]}<0.2$ and $5<\tau<6$~Gyr using Gaussian distributions centered on the median abundance. Here the total dispersion is computed as the standard deviation of the measured abundances. The measurement error term refers to the median error of the sample stars, while the intrinsic dispersion is derived with the MCMC modeling. 
The Figure shows that for O, Al, Si, Mn, Ni, and Nd, the intrinsic dispersion is remarkably smaller than the measurement error, while for other elements the intrinsic dispersion is comparable to the measurement error. The results suggest that abundances for most elements with atomic number smaller than Ni can be well reconstructed -- thus show little deviation -- from the abundance labels, which are in line with \citet{Ness2019}. However, for heavier elements such as Y, Zr, La, Ce, Sm and Eu, their abundances may exhibit substantial intrinsic dispersion, possibly due to distinct nucleosynthetic processes or environment variations in their production sites.  

\begin{figure}
\includegraphics[width=0.9\textwidth]{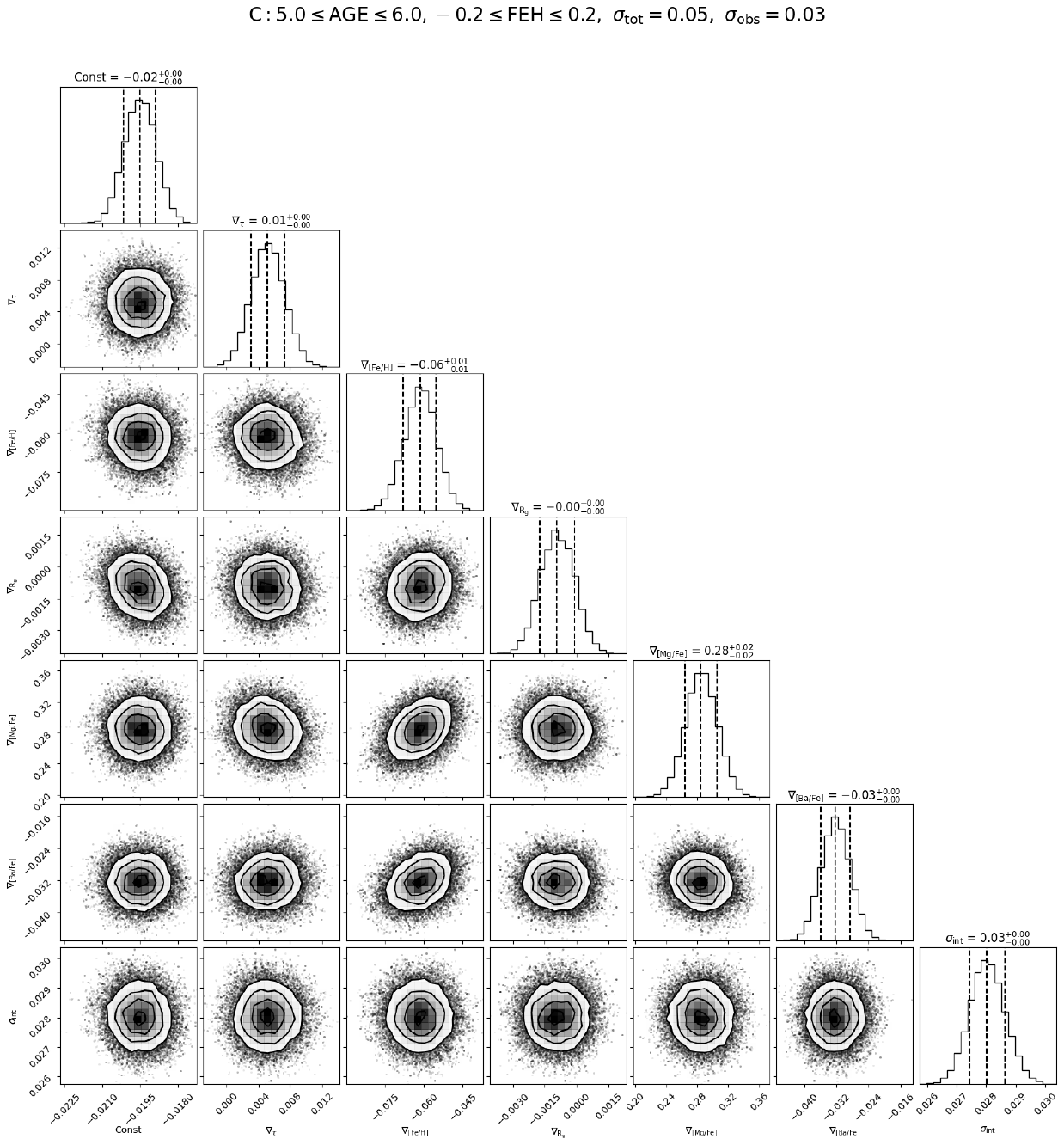}

\caption{An example of the posterior distributions obtained from the MCMC analysis for the intrinsic dispersion of [C/Fe] for stars in the bin centred $5<\tau<6$~Gyr and $-0.2<{\rm [Fe/H]}<0.2$~dex.} 
    \label{fig:fig4}
\end{figure}
\clearpage
\begin{figure*}
    \centering
    \includegraphics[width=\textwidth]{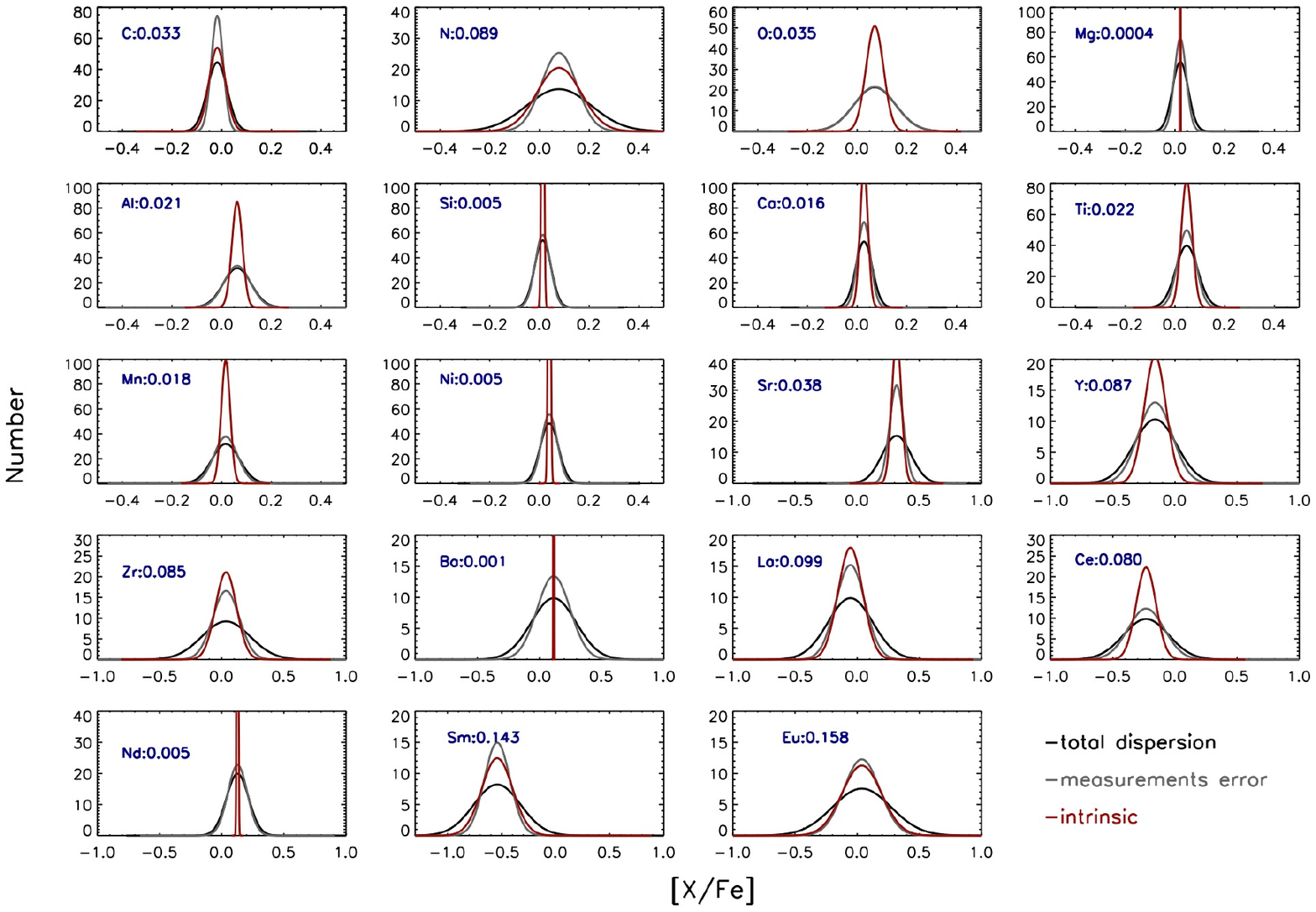}
    \caption{The distribution of total chemical abundances, measurement errors and intrinsic dispersions of 19 chemical element abundances for metallicity about 0\,dex and age about 5\,Gyr. }
\end{figure*}

\section{Results}

\subsection{Intrinsic abundance dispersion for stellar populations of different age and metallicity}

We characterize the intrinsic chemical abundance dispersion, as well as its relative value to the measurement error, $\sqrt{\sigma_{\mathrm{int}} / \sigma_{\mathrm{obs}}}$, for stellar populations of different stellar ages and metallicity. 

\begin{figure*}
 \centering
 \includegraphics[width=\textwidth]{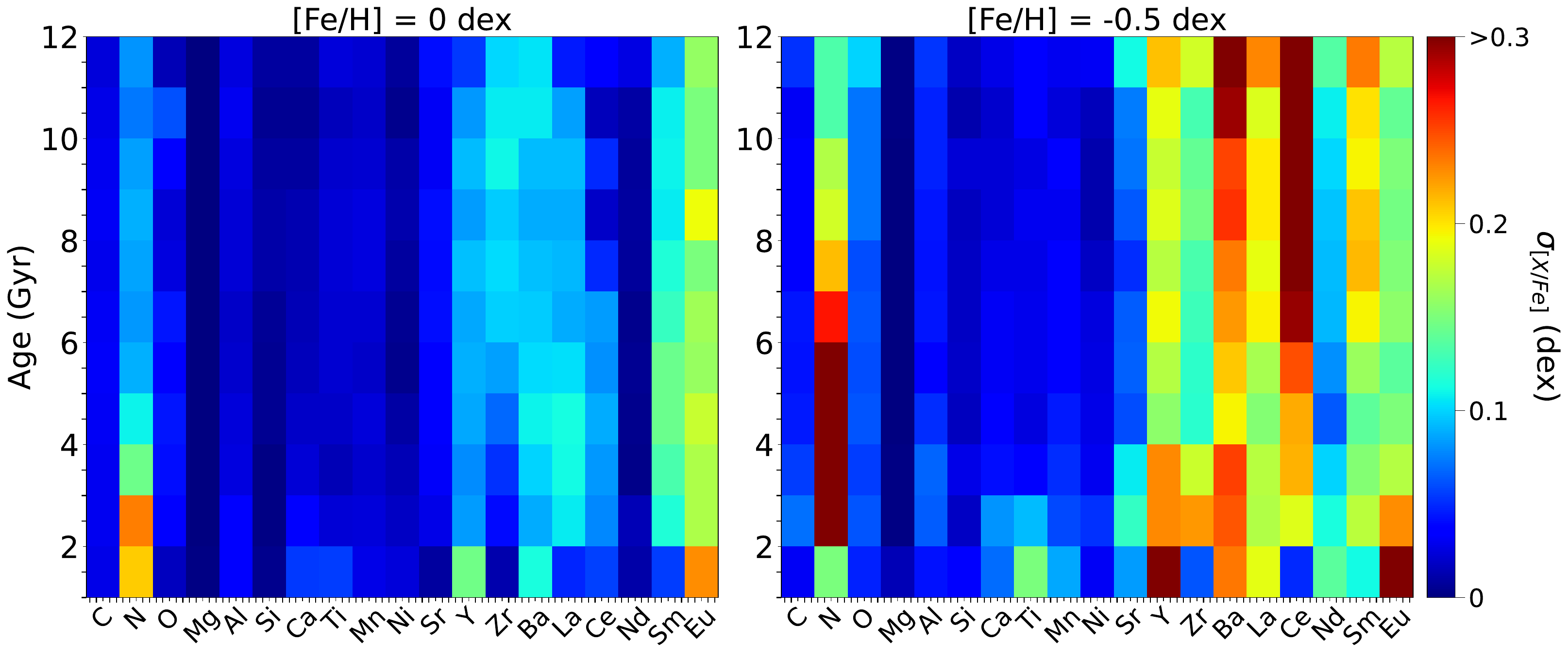}
 \includegraphics[width=\textwidth]{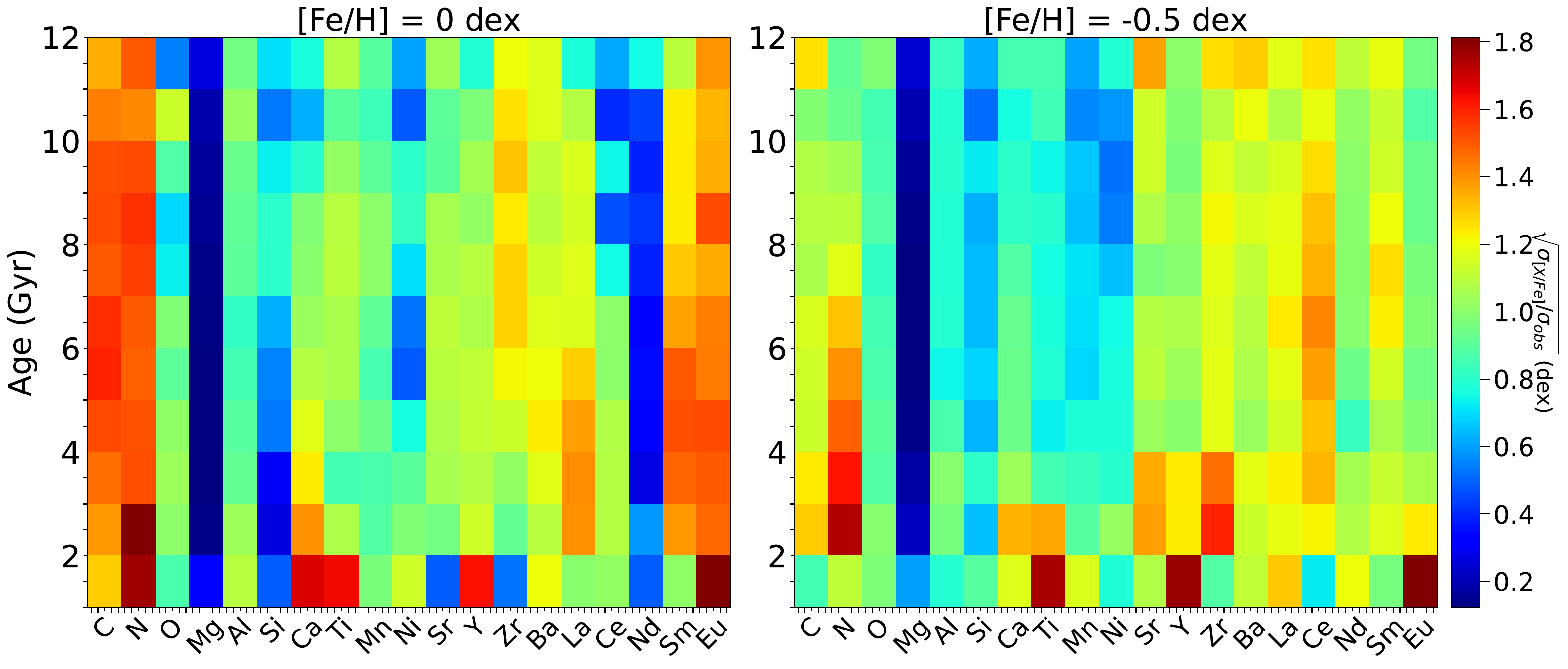}
 \caption{Intrinsic dispersion of elemental abundances as a function of stellar age. The top panel shows the intrinsic dispersion without [Ba/Fe] in the [X/Fe] model. The bottom panel illustrates the relative importance of the intrinsic dispersion compared to the observational uncertainties.}
    \label{fig:fig6}
 \end{figure*}

The upper panels of Figure\,6 illustrate the age dependence of intrinsic abundance dispersion for two metallicity regimes. Across both regimes, light, $\alpha$-, and iron-peak elements exhibit uniformly small intrinsic dispersions, typically below 0.05\,dex, with no significant variation as a function of age. In contrast, neutron-capture elements display substantially larger intrinsic dispersions, generally around $\sim$\,0.10\,dex near solar metallicity, and around $\sim$\,0.20\,dex for metal-poor regimes. Above them, two elements stand out as exceptions. N exhibits unusually large dispersions compared to other light elements, and its scatter declines mildly with increasing age. Nd, however, shows exceptionally small intrinsic dispersions among heavy elements. The reason for these exceptional results remains unclear.

A comparison of the left- and right- panels highlights the strong metallicity dependence of the intrinsic scatter. For light, $\alpha$ and iron-peak elements, the dispersions are  comparable between the metal-poor and solar metallicity regimes, suggesting the gas from which the stars formed are mixed to nearly the same extent for these different regimes. For neutron-capture elements, however, the metal-poor population exhibits substantially larger intrinsic dispersions, typically approaching $\sim$ 0.20 dex and clearly exceeding the values found near solar metallicity, suggesting the neutron-capture elements are less well mixed for the metal-poor population.

The lower panels of Figure\,6 illustrate the ratio between the intrinsic abundance scatter and the measurement errors across stellar age for the two metallicity regimes. Values above unity indicate that the intrinsic dispersion dominates over measurement errors in the scatter, while values below unity imply that the measured scatter is primarily error-driven.
Among the light, $\alpha$-, and iron-peak elements, C stands out with a noticeably higher ratio than other species in these groups, making it one of the few light elements that still retains meaningful discriminatory power for chemical tagging. N similarly exhibits a high ratio, consistent with its relatively large intrinsic dispersion, which dominates over measurement errors. In contrast, elements such as O, Al, Mn, and Ni show ratio well below unity. The exceptionally small residual dispersion in [Mg/Fe] primarily reflects its explicit inclusion as a conditioning variable in the model and should therefore not be directly compared with the dispersions of other elements.

For the neutron-capture species, the ratio remains systematically higher than for the lighter-element groups, reflecting their intrinsically larger star-to-star abundance variations, possibly driven by a mixture of s- and r-process channels. 
Nd, however, shows a distinct behavior compared to other heavy elements. In contrast to the general trend, its ratio is essentially zero for the solar metallicity bin but becomes significant for the metal-poor regime. 
This cause of such a dispersion pattern for Nd is unclear. 

Figure\,7 (top panel) shows that the three age populations exhibit remarkably similar trends, indicating that metallicity—rather than age—is the dominant factor shaping intrinsic abundance dispersion. For the light, $\alpha$-, and iron-peak elements, the intrinsic dispersions remain uniformly small (generally below 0.05 dex) and display only weak dependence on [Fe/H] across all age bins. 
In contrast, the neutron-capture elements show both larger and strongly metallicity-dependent intrinsic dispersions. At lower metallicities ([Fe/H] $\lesssim$ $-0.4$), heavy elements exhibit significantly enhanced scatter, often exceeding 0.1 dex, whereas their dispersions systematically decline toward solar metallicity. For neutron-capture elements, the s-process–dominated species (e.g., Y, La, Ce, Ba) display the strongest metallicity gradients. In contrast, the r-process–dominated elements (e.g., Sm, Eu) show little dependence on [Fe/H]. 

This overall trend highlights the fundamentally different behavior of the abundance scatter in light versus neutron-capture elements.
For the light, $\alpha$, and iron-peak species, the low intrinsic dispersions agree well with previous findings from \citet{Ness2019} and other chemical-tagging studies \citep[e.g.]{Manea2022,Manea2024,Sinha2024}. These results consistently suggest that light and iron-peak elements exhibit nearly homogeneous abundance patterns in the Galactic disk and thus provide limited discriminatory power for chemical tagging. In contrast, the neutron-capture elements exhibit systematically higher intrinsic dispersion, reflecting the far richer diversity and stochasticity of the nucleosynthetic processes that govern their production.  This elevated intrinsic scatter for neutron-capture elements have been also reported by previous studies, such as \citet{Matsuno2021,Griffith2021,Sinha2024}, and suggest that these elements carry the strongest chemical “signal” for  chemical tagging.

\begin{figure*}
 \centering
 \includegraphics[width=\textwidth]{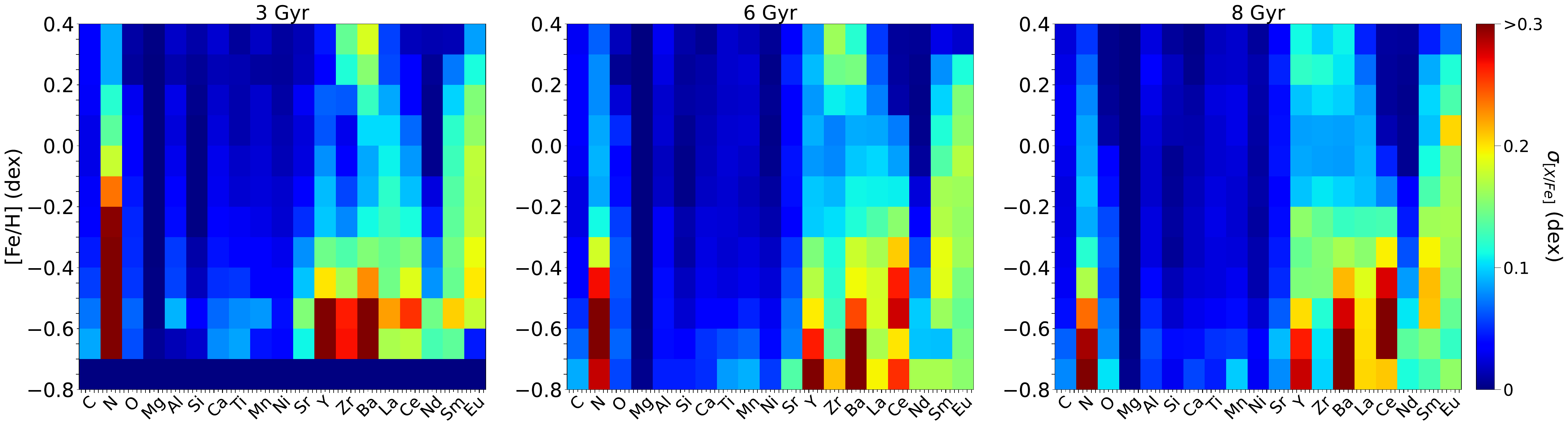}
 \includegraphics[width=\textwidth]{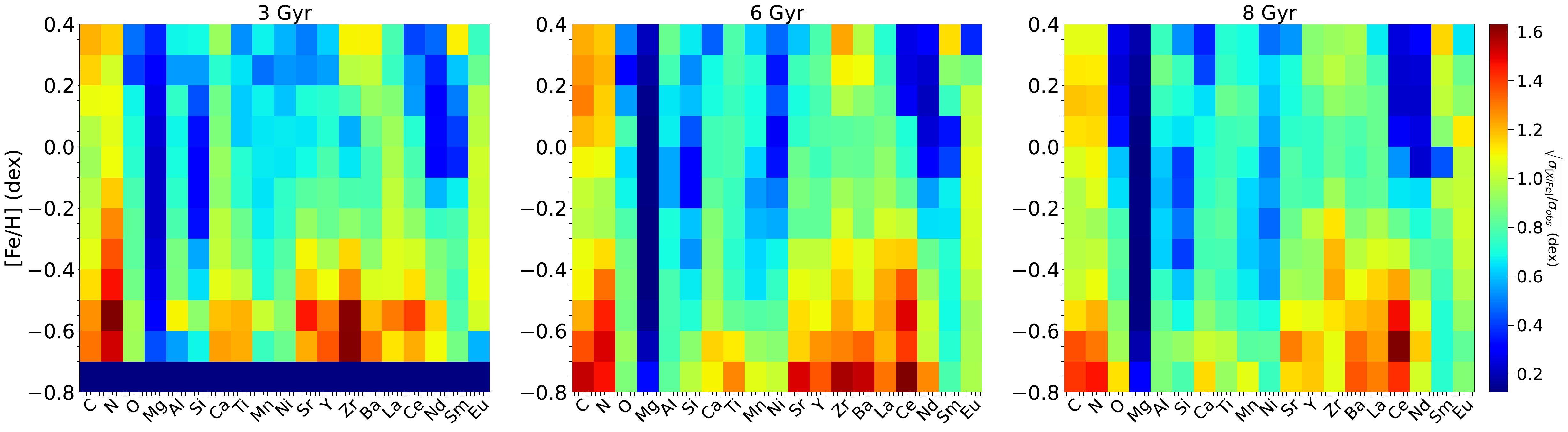}
 \caption{Intrinsic dispersion of elemental abundances as a function of metallicity. The top panel shows the intrinsic dispersion without [Ba/Fe] in the [X/Fe] model. The bottom panel illustrates the relative importance of the intrinsic dispersion compared to the observational uncertainties.}
    \label{fig:fig7}
 \end{figure*}

\subsection{Comparison with M67}

To assess the robustness of our intrinsic abundance-dispersion measurements, we analyze the open cluster M67, a classical benchmark system that is widely regarded as chemically homogeneous. We selected approximately 100 main-sequence turn-off (MSTO) stars from the LAMOST DR9 value-added catalog, excluding known binaries to minimize the influence of multiplicity. For these stars, we measured intrinsic dispersions for 19 elemental abundance ratios spanning light, $\alpha$-, iron-peak, and neutron-capture elements.

Figure\,8 (grey dotted line) shows the inferred intrinsic dispersions and their $1\sigma$ uncertainties. As expected for a cluster generally regarded as chemically homogeneous, M67 exhibits very small abundance dispersions: all elements lie below 0.1 dex, with light, $\alpha$-, and iron-peak elements typically below 0.05 dex. Neutron-capture elements show only slightly larger dispersions, by $\sim$0.02--0.03 dex.

For validation, Figure,8 also includes the intrinsic dispersions reported by \citet{Bovy2016}. Our results agree well with their high-resolution benchmark values for most light, $\alpha$-, and iron-peak elements (with the exception of N), providing independent support for the intrinsic-dispersion measurements derived from low-resolution spectroscopy.

Compared with M67, field subgiants of similar metallicity generally exhibit larger intrinsic abundance dispersions, particularly for neutron-capture elements. By contrast, Al, Mg, and Nd show comparable, and in a few cases marginally larger, dispersions in M67 than in the field sample. These individual differences should be interpreted with caution because the M67 sample is dominated by MSTO stars whereas the field sample consists of subgiants. Previous studies have shown that atomic diffusion can produce measurable abundance differences between these evolutionary stages when absolute abundances ([X/H]) are considered \citep[e.g.][]{Liu2019}. In the present work, however, we analyse abundance ratios relative to iron ([X/Fe]), for which diffusion effects are expected to be substantially reduced because both iron and the element of interest are affected in a similar manner. Although some residual evolutionary effects may remain, they are unlikely to dominate the comparison.

We therefore regard M67 primarily as an empirical benchmark for validating our methodology rather than as a strict reference for the absolute level of abundance dispersion. Nevertheless, the systematically larger dispersions observed for neutron-capture elements in the field-star sample indicate that these elements preserve signatures of more diverse chemical-enrichment histories than are present within a single open cluster. In contrast, the similarly small dispersions found for most light and $\alpha$ elements in both M67 and the field suggest that these species were efficiently mixed before star formation and therefore provide relatively limited discriminatory power for chemical tagging. These results are broadly consistent with previous high-resolution studies \citep[e.g.][]{Liu2016,Sinha2024} and reinforce the view that neutron-capture elements constitute the most informative chemical tracers for distinguishing co-natal stellar populations. The overall agreement with independent high-resolution measurements further demonstrates that abundance dispersions derived from LAMOST low-resolution spectra can reliably recover the principal chemical signatures associated with Galactic chemical evolution.

\begin{figure*}
\centering
\includegraphics[width=\textwidth]{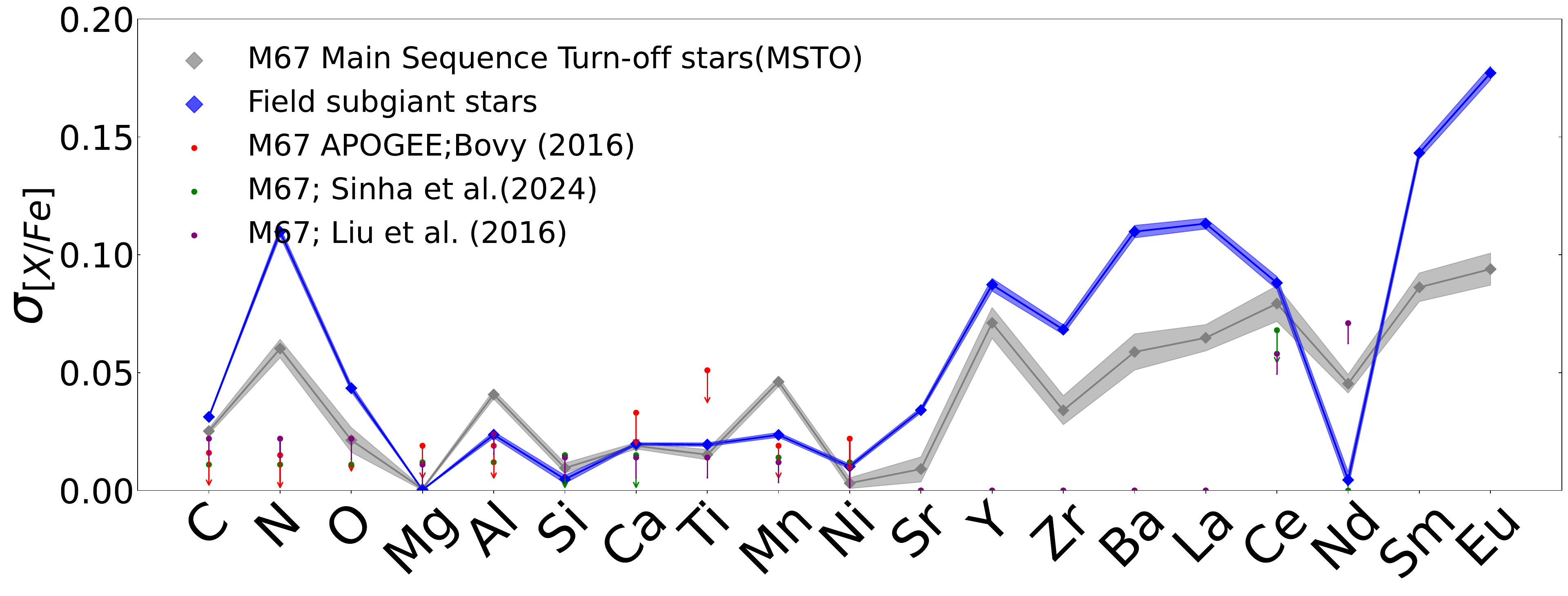}
\caption{Intrinsic abundance dispersions for 19 elements measured for field stars and for the open cluster M67. Blue diamonds show the intrinsic dispersions inferred for field stars with ages between 4 and 5\,Gyr and metallicities in the range $-0.2 < \mathrm{[Fe/H]} < 0.2$\,dex. Gray diamonds indicate the corresponding results obtained from a model that does not take [Ba/Fe] into account. Shaded regions represent the $1\sigma$ uncertainties of the inferred intrinsic dispersions. For comparison, red circles denote the upper limits on the intrinsic abundance scatter of M67 derived from APOGEE data by \citet{Bovy2016}, green circles show the upper limits reported by \citet{Sinha2024}, and purple circles indicate the measurements from \citet{Liu2016}.}  \label{fig:fig8}
\end{figure*}

\subsection{Dispersion associated with [Ba/Fe]}

Neutron-capture elements play a central role in tracing delayed and environment-dependent nucleosynthetic processes in stellar populations. Among these, the s-process products from low- and intermediate-mass AGB stars provide a key probe of long-timescale enrichment. Because Ba is predominantly produced by the main s-process, we adopt [Ba/Fe] as the primary reference variable to isolate the contribution of delayed AGB enrichment when modeling intrinsic abundance dispersion. 

Figures\,9, 10 and 11 show the newly derived intrinsic dispersions after applying Ba as a reference label of the abundance model (Equation~4), which are counterparts of Figures\,6, 7 and 8, respectively. The patterns shown in these figures remain largely unchanged compared to those that do not consider the Ba dependence (Figures~6, 7, and 8), except for the fact that the [Ba/Fe] dispersion becomes zero, as expected. For $\alpha-$elements and iron-peak elements, the results are expected, as these elements are synthesized in different channels to Ba. However, the small change in s-process elements other than Ba represents a new, surprising result. This suggests that their abundance variations cannot be described solely by Ba enrichment, and that the neutron-capture elements in our sample stars may have had diverse nucleosynthetic origins. Notably, this is the case even for M67, as the member stars exhibit substantially large abundance scatter ($\simeq0.08$~dex) for the neutron-capture elements. 

Figure\,12 shows the comparison of intrinsic scatter derived without and with [Ba/Fe] in the [X/Fe] modelling (i.e., Equations~3 and 4, respectively), for both the solar-metallicity stellar population and the stellar population of $-0.6<{\rm [Fe/H]}<-0.4$ and $5<\tau<6$~Gyr. It illustrates that while the inclusion of Ba in the modelling had little effect on the abundance dispersion for the case of solar-metallicity population, it does reduce the dispersion of [Y/Fe], [La/Fe], and [Ce/Fe] slightly for the relatively metal-poor population. This could be evidence for early and coherent s-process contributions from metal-poor AGB stars \citep{Bisterzo2012,Placco2013}. Meanwhile, abundance dispersions for Nd, Sm, and Eu remain unaffected, reflecting the fundamentally stochastic nature of the r-process, produced in rare, spatially inhomogeneous events such as neutron-star mergers or magneto-rotational supernovae \citep[e.g.][]{Sneden2008, Cote2019, Kobayashi2020, Cowan2021}. These metal-poor results agree with previous conclusions that r-process abundances retain large intrinsic scatter and provide strong chemical-tagging leverage even at moderate spectral resolution \citep{Manea2024, Sinha2024}.

\begin{figure*}
 \centering
 \includegraphics[width=\textwidth]{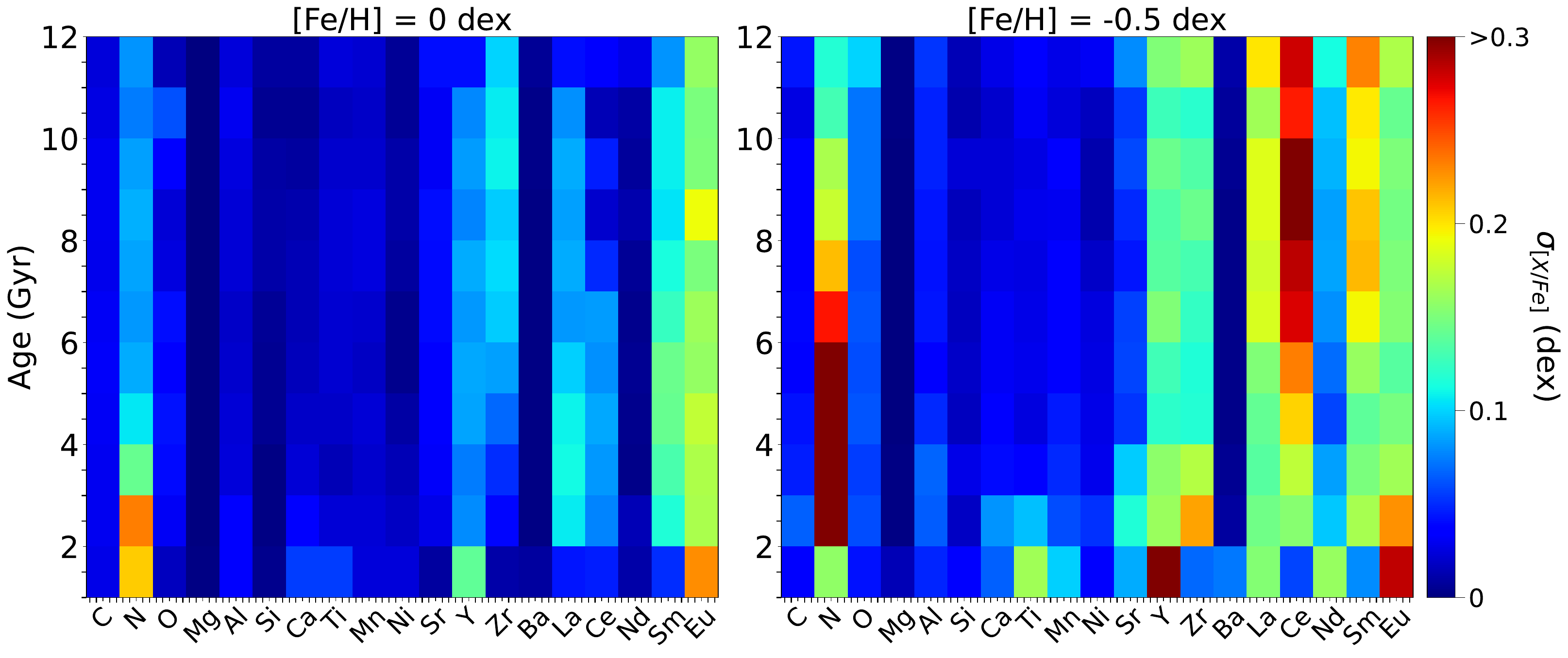}
 \includegraphics[width=\textwidth]{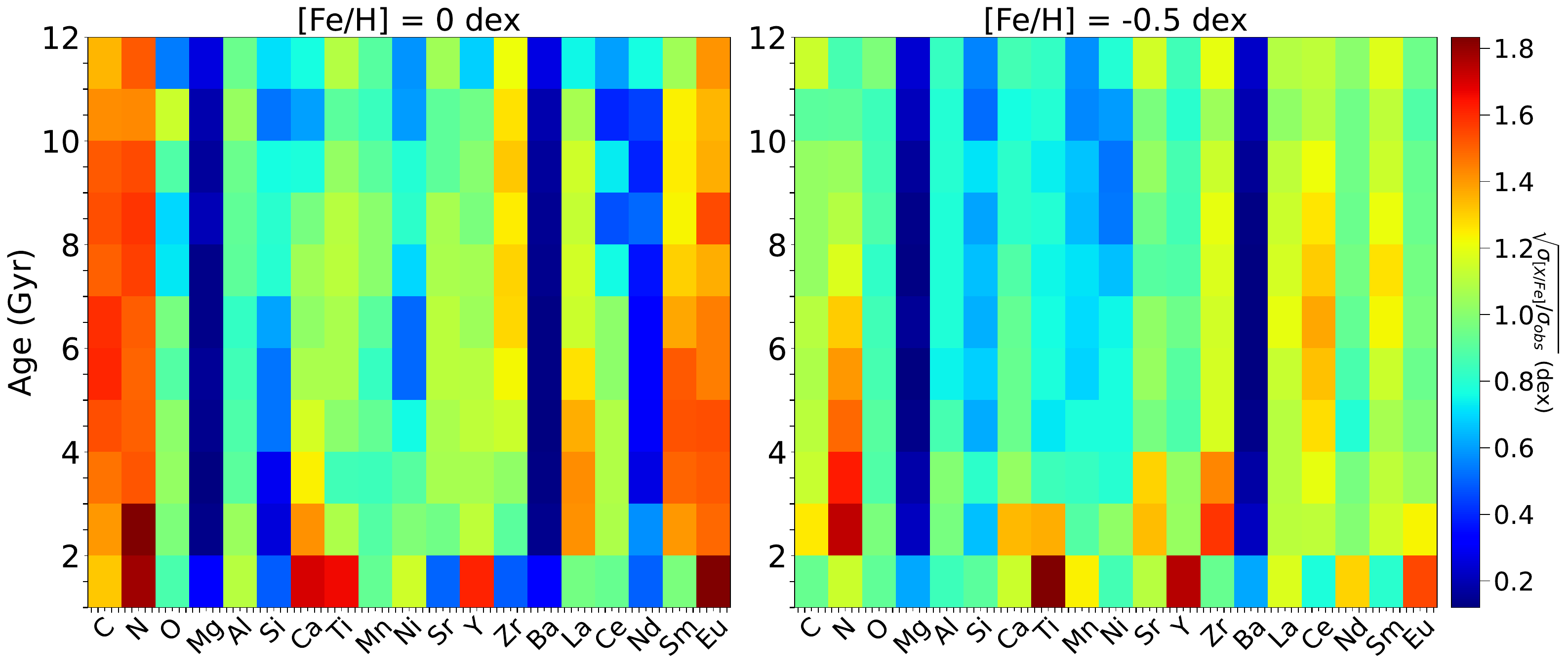}
 \caption{Same as Figure~5, but including [Ba/Fe] in the model.}
    \label{fig:fig9}
 \end{figure*}

 \begin{figure*}
 \centering
 \includegraphics[width=\textwidth]{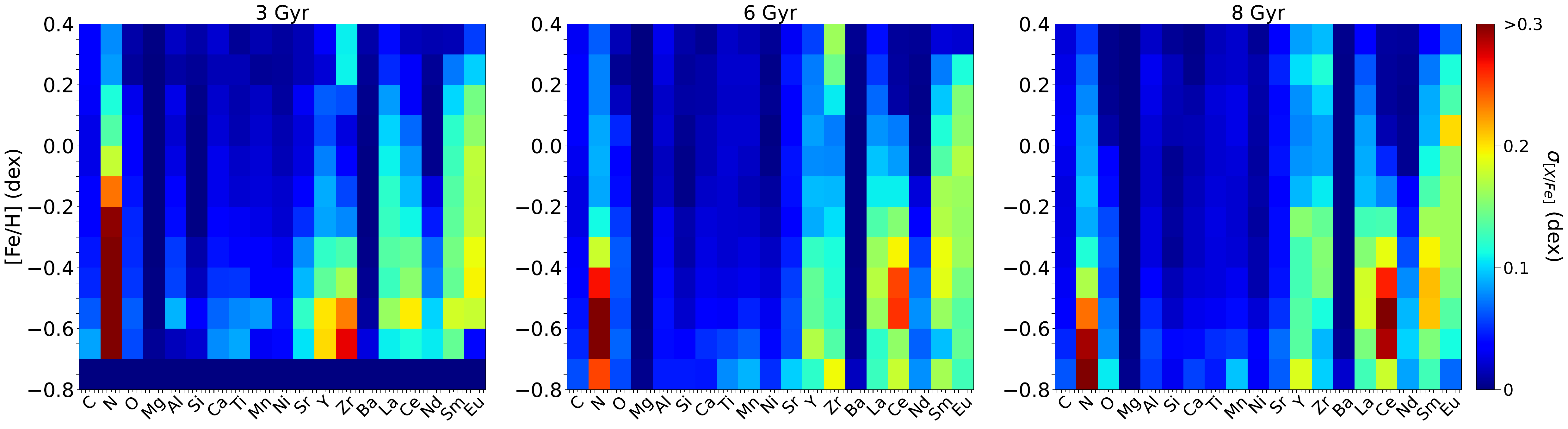}
 \includegraphics[width=\textwidth]{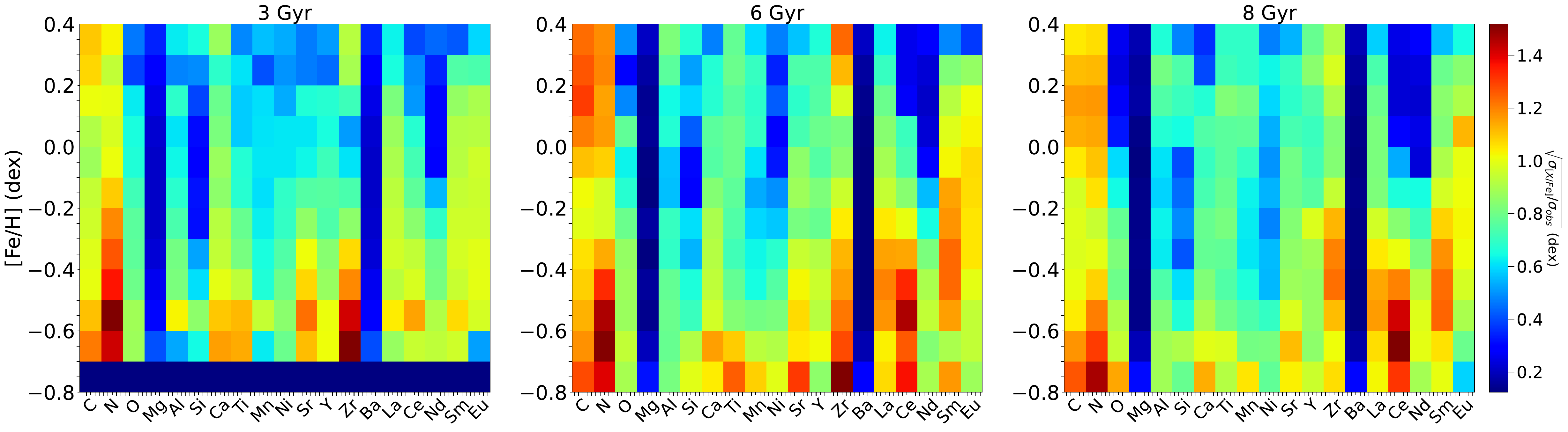}
 \caption{Same as Figure\,6, but with [Ba/Fe] in the [X/Fe] model.}
    \label{fig:fig10}
 \end{figure*}

\begin{figure*}
\centering
\includegraphics[width=\textwidth]{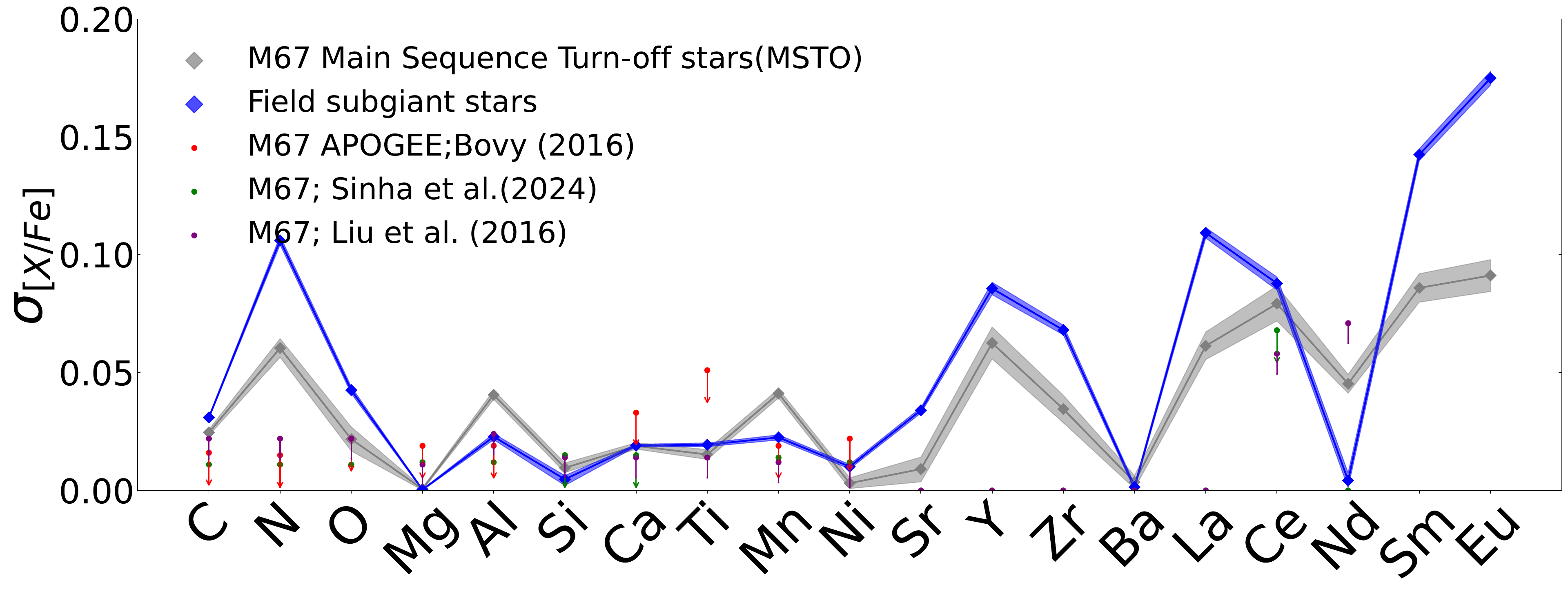}
\caption{Same as Figure\,7, but with [Ba/Fe] in the [X/Fe] model.}  \label{fig:fig11}
\end{figure*}

\begin{figure*}
\centering
\includegraphics[width=\textwidth]{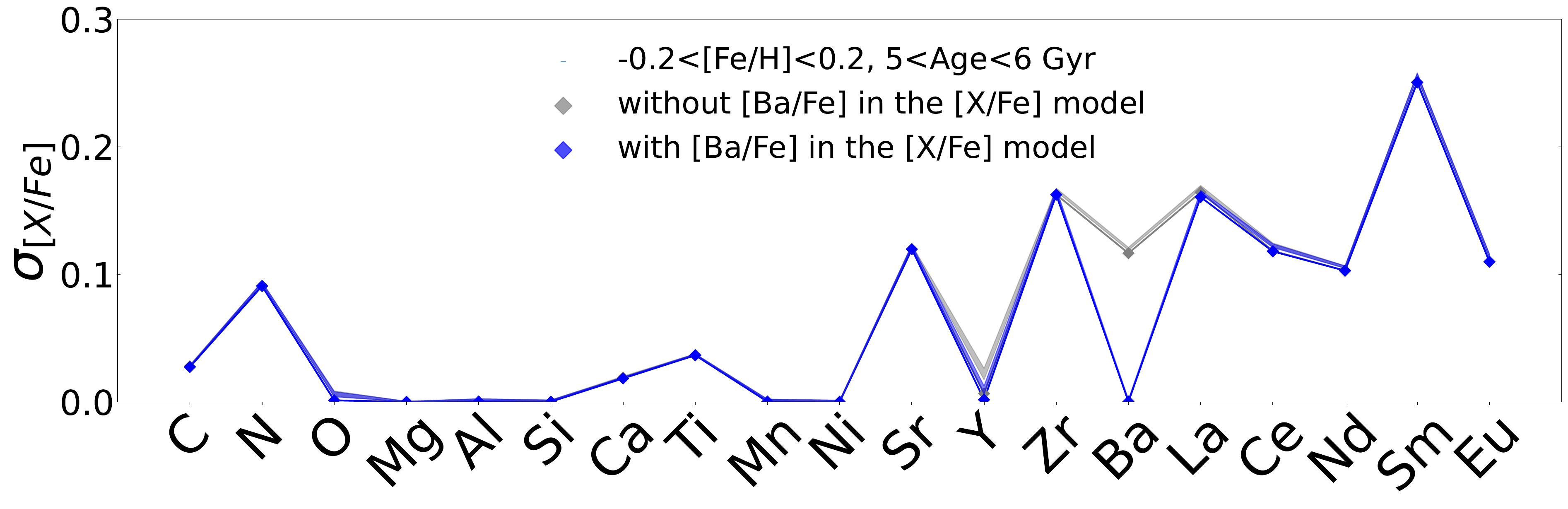}
\includegraphics[width=\textwidth]{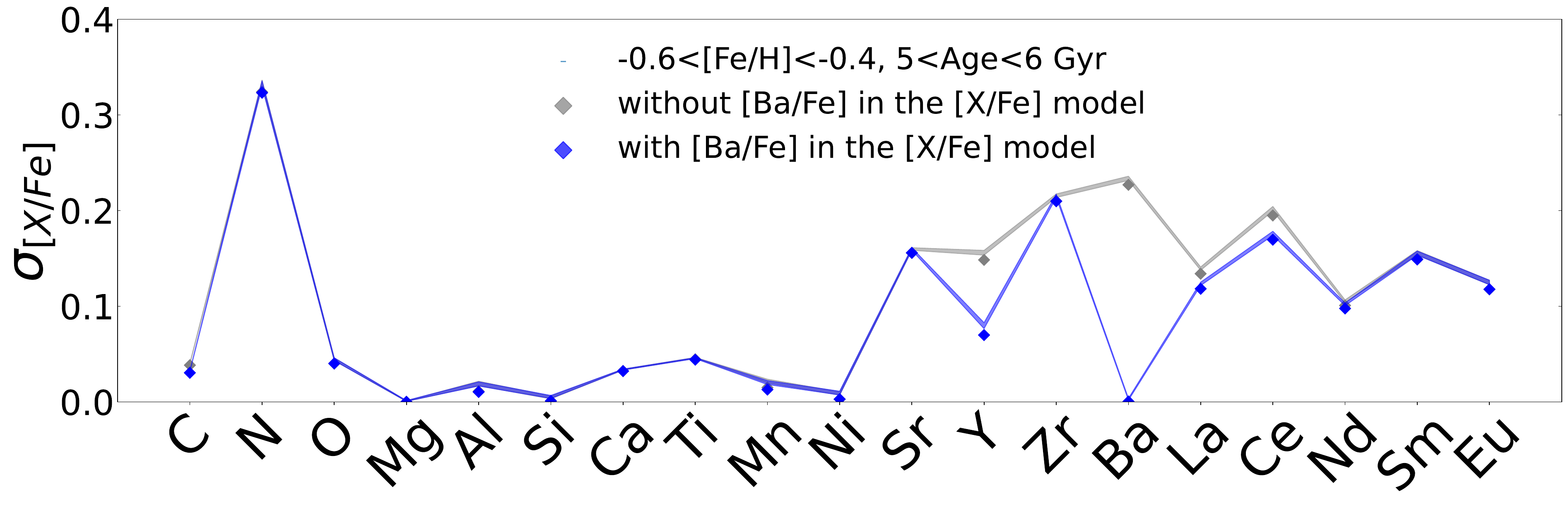}
\caption{Upper panel: Intrinsic abundance dispersions for 19 elements measured for metal-rich stars with ages between 5 and 6\,Gyr. Gray diamonds show the results obtained from the [X/Fe] model without including [Ba/Fe], while blue diamonds indicate the corresponding dispersions inferred from the model that includes [Ba/Fe].
Lower panel: Same as the upper panel, but for metal-poor stars with ages between 5 and 6\,Gyr.}  \label{fig:fig12}
\end{figure*}

\section{Discussion and Implication}

Our results demonstrate that intrinsic chemical abundance dispersion is a fundamental property of stellar populations -- even in systems traditionally regarded as chemically homogeneous, such as open clusters like M67. While M67 shows mean elemental abundances consistent with previous high-resolution studies \citep[e.g.,][]{Bovy2016, Poovelil2020, Sinha2024}, we find that the element-level intrinsic scatter for several species, including C, Si, Ca, Ti, and Ni, are comparable to field stars of similar metallicity. Furthermore, the intrinsic abundance scatter among individual stars is larger than 0.05~dex for almost all the neutron-capture elements. These results suggest that measurable chemical inhomogeneities may be present not only in the Galactic scale, but also even within small-scale, coeval open clusters. 
This highlights limitations of assuming perfect homogeneity and limits the ability to cleanly separate cluster stars from the field with chemical tagging.

Incorporating [Ba/Fe]—a representative tracer of s-process nucleosynthesis—into our modeling leads to only a minor reduction in the intrinsic scatter of other s-process elements including Y, La, and Ce. For other neutron-capture elements, particularly those with dominant r-process contributions (Sm, Eu), the inclusion of [Ba/Fe] in the modelling induces negligible changes in their inferred dispersion. These results reveal a more complex and possibly stochastic enrichment history for the individual stars, both for field stars and cluster members.

These findings raise important caveats for strict chemical tagging. The relatively small intrinsic scatter in light, $\alpha$-, and iron-peak elements limits their ability to discriminate between co-natal and non-co-natal stars at solar metallicity. Even with precise abundance measurements, the chemical similarity among unrelated stars leads to significant overlap in abundance space. Although the inclusion of neutron-capture elements offers additional power to identify co-natal stars, as has been suggested in previous studies \citep[e.g.,][]{Manea2024, Sinha2024} and been confirmed in the current work, the substantial intrinsic scatter in coeval cluster members impose challenges for a comprehensive exercise of chemical tagging.

To improve the accuracy of chemical tagging and better understand intrinsic abundance dispersion, future research should focus on:
\begin{itemize}
\item Examining and validating the current results using better, more precise abundance measurements from high-resolution spectroscopic surveys;
\item Expanding the sample size of open clusters with diverse ages and metallicities to accurately map baseline chemical variations;
\item Investigating stellar populations across different evolutionary phases (e.g., main-sequence turnoff, subgiants, red giants) to assess the role of stellar evolution on chemical dispersion;
\item Enhancing the accuracy of atomic data and spectral line lists for neutron-capture elements to reduce measurement uncertainties and improve model reliability.
\end{itemize}

\section{Conclusions}

Stellar abundance dispersions encode important information about the chemical enrichment history of the interstellar medium and the degree of mixing within star-forming environments. In this work, we have measured the intrinsic abundance dispersions of 19 elements using a large sample of low-$\alpha$ subgiant stars with precise age estimates from Gaia--LAMOST. By constructing a multivariate model that accounts for the dependence of elemental abundances on stellar age, metallicity, [Mg/Fe], and [Ba/Fe], we isolate the residual star-to-star abundance scatter and investigate its variation across different stellar populations.

Our principal conclusions are as follows:

\begin{itemize}

\item The intrinsic abundance dispersions of light elements, $\alpha$ elements, and iron-peak elements are generally very small, typically below 0.05 dex, whereas neutron-capture elements exhibit substantially larger dispersions, often exceeding 0.1 dex. This striking contrast reflects their different nucleosynthetic origins and enrichment timescales, indicating that heavy elements preserve information about localized chemical enrichment that is largely erased for lighter species.

\item The intrinsic abundance dispersions show little systematic dependence on stellar age but exhibit a clear dependence on metallicity. Relatively metal-poor populations consistently display larger dispersions, suggesting that star-forming environments in the early Galactic disc were chemically less homogeneous than those at later times.

\item The inferred intrinsic abundance dispersions in the benchmark open cluster M67 are non-zero under our modelling framework. For several $\alpha$- and iron-peak elements, the inferred dispersions are comparable to those measured in field stars, whereas neutron-capture elements remain systematically more homogeneous than the field population, although they are not perfectly uniform. These results indicate that the assumption of perfect chemical homogeneity should be relaxed, while simultaneously demonstrating that neutron-capture elements provide substantially greater discriminatory power than light elements for identifying co-natal stellar populations.

\end{itemize}

Taken together, our results suggest that measurable intrinsic abundance dispersion is a common property of stellar populations rather than an exceptional phenomenon. The presence of non-zero abundance scatter, even within a well-studied open cluster such as M67, implies that classical chemical tagging based on the assumption of perfect homogeneity requires refinement. However, the much larger star-to-star variations exhibited by neutron-capture elements indicate that these species retain unique signatures of localized nucleosynthetic enrichment and therefore remain the most powerful chemical tracers for reconstructing stellar birth sites and Galactic assembly history.

Finally, this work demonstrates that, after careful calibration and validation against high-resolution spectroscopic surveys, low-resolution spectroscopy can robustly recover intrinsic abundance dispersions at the $\lesssim0.05$ dex level for most light elements. Combined with precise stellar ages and large statistical samples, such data provide a powerful and scalable framework for investigating Galactic chemical evolution and for advancing next-generation chemical-tagging studies involving millions of stars.

\section*{Acknowledgements}
We acknowledge financial support from the National Natural Science Foundation of China (NSFC) under grant No.12588202, National Key R\&D Program of China No.2023YFE0107800, No.2022YFE0504200, No.2024YFA1611900, Strategic Priority Research Program of Chinese Academy of Sciences, grant No.1160102. We also acknowledge support from the China Manned Space Program with grant No.CMS-CSST-2025-A12. M.X. acknowledges financial support from NSFC Grant No.2022000083. and National Key R\&D Program of China Grant No.2022YFF0504200. 

This work made use of the data from LAMOST (Large Sky Area Multi-Object Fiber Spectroscopic Telescope, also known as the Guoshoujing Telescope) (https://cstr.cn/31118.02.LAMOST ). LAMOST is a Chinese national mega-science facility, operated by National Astronomical Observatories, Chinese Academy of Sciences.

%%%%%%%%%%%%%%%%%%%%%%%%%%%%%%%%%%%%%%%%%%%%%%%%%%
\section*{Data Availability}

The data underlying this article will be shared on reasonable request to the corresponding author.
%The inclusion of a Data Availability Statement is a requirement for articles published in MNRAS. Data Availability Statements provide a standardised format for readers to understand the availability of data underlying the research results described in the article. The statement may refer to original data generated in the course of the study or to third-party data analysed in the article. The statement should describe and provide means of access, where possible, by linking to the data or providing the required accession numbers for the relevant databases or DOIs.



\bibliographystyle{mnras}
\bibliography{references} % if your bibtex file is called example.bib

%\begin{thebibliography}{99}
%\end{thebibliography}

%%%%%%%%%%%%%%%%%%%%%%%%%%%%%%%%%%%%%%%%%%%%%%%%%%

%%%%%%%%%%%%%%%%% APPENDICES %%%%%%%%%%%%%%%%%%%%%

\appendix

\section{Validation against GALAH DR4}

The intrinsic abundance dispersions derived in this work rely on the accuracy and stability of the underlying DD-Payne abundance measurements. Although the DD-Payne labels have been extensively validated in \citet{Xiang2019,Zhang2025}, the present study focuses on a specific sample of low-$\alpha$ subgiants occupying a relatively narrow region of parameter space. It is therefore important to verify that the adopted abundances do not exhibit significant systematic offsets or parameter-dependent trends within this sample.

To this end, we cross-match our final sample with GALAH DR4 \citep{Buder2025} and compare both stellar parameters and elemental abundances for the overlapping stars. In addition, we examine the abundance residuals as functions of effective temperature and metallicity in order to identify possible systematic trends that could artificially contribute to the inferred intrinsic abundance dispersions.

Figure~A1 compares the stellar parameters derived by DD-Payne and GALAH DR4. The two data sets are in excellent agreement, with only small median offsets and robust scatters consistent with the expected uncertainties of low-resolution spectroscopy.

The elemental abundance comparisons are presented in Figure~A2. Most light, $\alpha$-, and iron-peak elements show good agreement with the GALAH abundance scale, with median offsets close to zero and dispersions comparable to the quoted measurement uncertainties. Larger offsets are found for several neutron-capture elements, particularly Ba and Eu, reflecting the greater difficulty of measuring these species from low-resolution spectra.

To further investigate possible systematic effects, Figures~A3 and A4 show the abundance residuals,$\Delta[X/{\rm Fe}] = [X/{\rm Fe}]_{\rm DD\text{-}Payne} - [X/{\rm Fe}]_{\rm GALAH}$, as functions of effective temperature and metallicity, respectively. The residuals generally vary smoothly behaviour across the parameter space occupied by our low-$\alpha$ subgiant sample. Magnesium, silicon, and titanium remain nearly constant with both $T_{\rm eff}$ and [Fe/H], indicating excellent consistency between the two surveys. Carbon, oxygen, calcium, and barium exhibit mild systematic trends with effective temperature, while carbon shows a gradual decrease and calcium and manganese display weak positive trends with metallicity. Nickel exhibits only a slight metallicity dependence, and yttrium remains largely stable over the sampled parameter range. The largest excursions are primarily confined to the edges of the parameter space, where the number of overlapping stars decreases substantially and the uncertainties become correspondingly larger.

Importantly, these parameter-dependent variations are smooth and generally limited to amplitudes of order $\sim0.1$--$0.2$ dex over the full parameter range. They therefore cannot account for the substantially larger intrinsic abundance dispersions observed for several neutron-capture elements in the main analysis. Moreover, no abrupt discontinuities or strong systematic structures are present that would artificially generate the measured abundance scatter. These comparisons provide independent evidence that the DD-Payne abundance scale is sufficiently stable for studies of intrinsic chemical dispersion and that the principal conclusions of this work are unlikely to be driven by temperature- or metallicity-dependent calibration effects.

\begin{figure*}
    \centering
    \includegraphics[width=\textwidth]{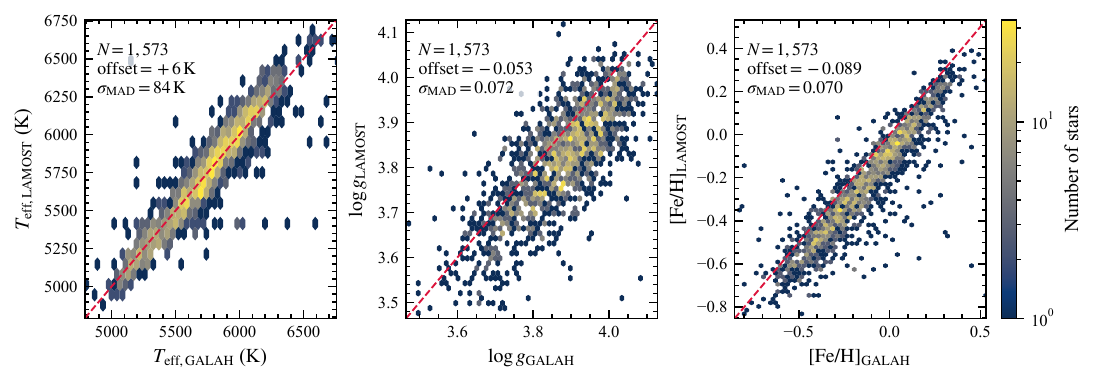}
    \caption{Comparison of stellar parameters between the DD-Payne measurements and GALAH DR4 for the overlapping subgiant sample. From left to right, the panels show comparisons of $T_{\rm eff}$, $\log g$, and [Fe/H]. The dashed lines indicate one-to-one relations. The median offsets and robust scatters ($\sigma_{\rm MAD}$) are listed in each panel.}
\end{figure*}

\begin{figure*}
    \centering
    \includegraphics[width=\textwidth]{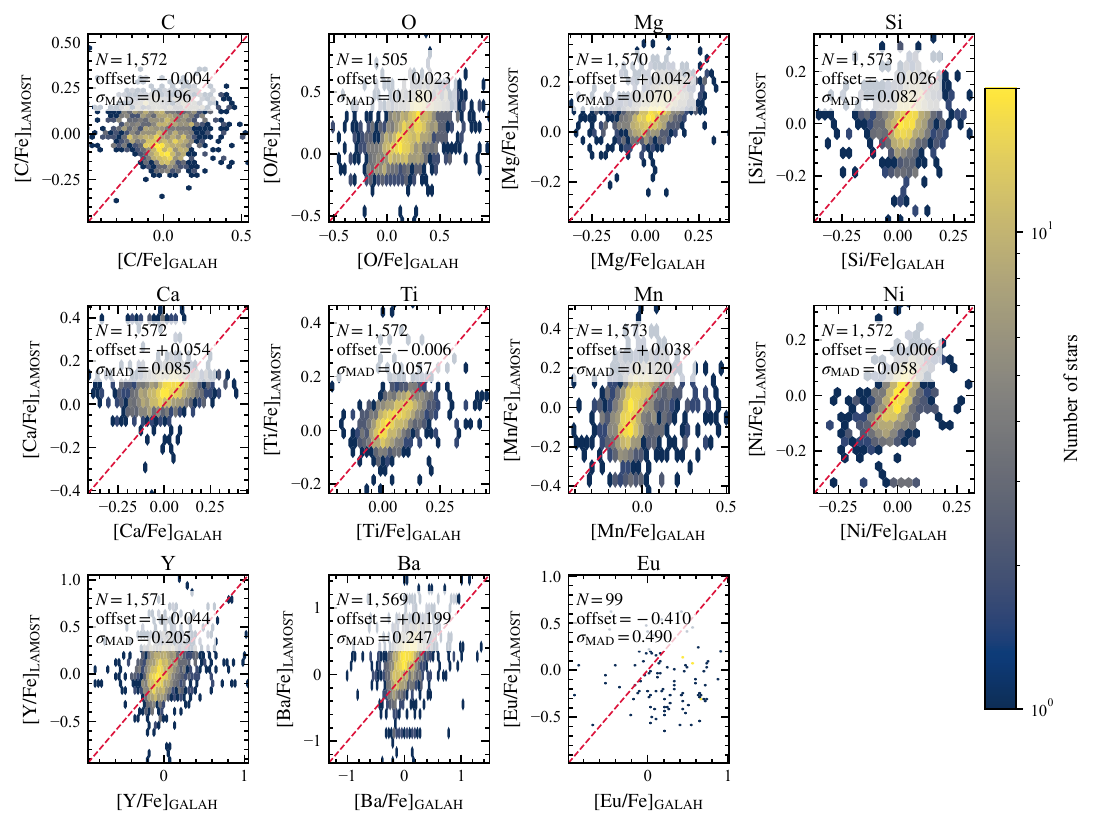}
    \caption{Comparison of elemental abundances between DD-Payne and GALAH DR4. Each panel shows one abundance ratio measured by the two surveys. The dashed line indicates the one-to-one relation. The number of stars, median abundance offset, and robust scatter ($\sigma_{\rm MAD}$) are reported in each panel.}
\end{figure*}

\begin{figure*}
    \centering
    \includegraphics[width=\textwidth]{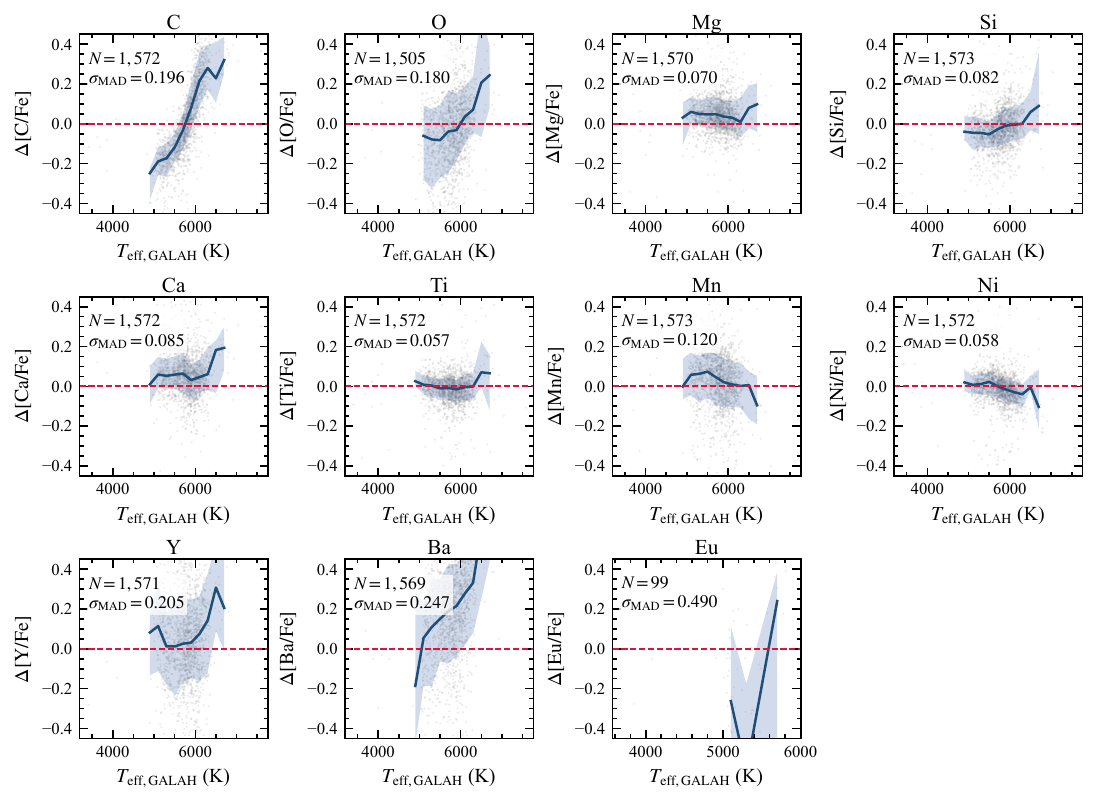}
    \caption{Abundance residuals, defined as $\Delta[X/{\rm Fe}] = [X/{\rm Fe}]{\rm DD\text{-}Payne} - [X/{\rm Fe}]{\rm GALAH}$, as a function of effective temperature. The solid curves show the running median, while the shaded regions indicate the 16th--84th percentile ranges. Most elements exhibit only weak temperature-dependent systematics across the parameter range of the final sample.}
\end{figure*}

\begin{figure*}
    \centering
    \includegraphics[width=\textwidth]{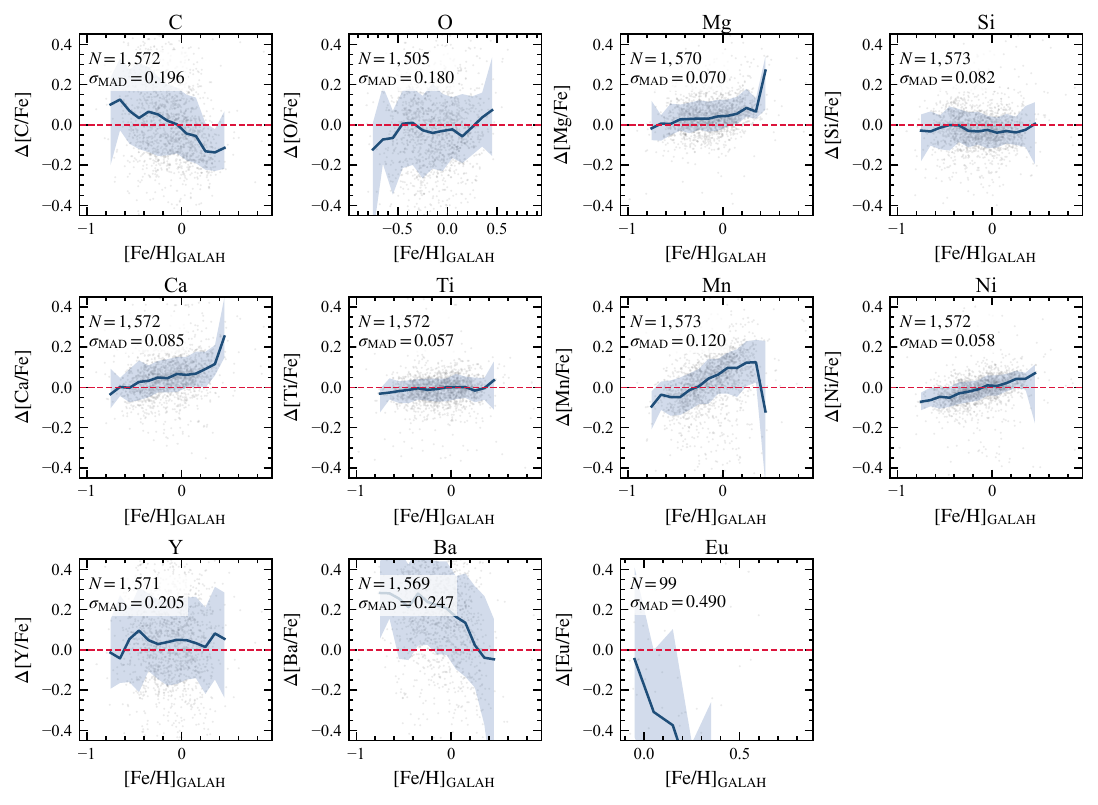}
    \caption{Abundance residuals, defined as $\Delta[X/{\rm Fe}] = [X/{\rm Fe}]{\rm DD\text{-}Payne} - [X/{\rm Fe}]{\rm GALAH}$, as a function of metallicity. The solid curves show the running median, while the shaded regions indicate the 16th--84th percentile ranges. No strong metallicity-dependent biases are observed for most elements within the metallicity range relevant to this study.}
\end{figure*}

\section{Validation against APOGEE DR19}

As an independent validation, we further compare the DD-Payne abundances with APOGEE DR19 \citep{Meszaros2025} which provides high-resolution near-infrared spectroscopy and abundance measurements derived using different spectral features and analysis methodologies from those employed by GALAH. Although the overlap with our low-$\alpha$ subgiant sample is smaller than that with GALAH DR4, the APOGEE comparison offers an important external assessment of the reliability of the abundance scale adopted in this work.

Figure~B1 presents the comparison of stellar parameters between DD-Payne and APOGEE DR19. The inferred effective temperatures, surface gravities, and metallicities show good overall agreement, with only modest systematic offsets and robust scatters consistent with expectations for low-resolution spectroscopic determinations.

The elemental abundance comparisons are shown in Figure~B2. Despite the different wavelength coverage and independent analysis pipelines, the DD-Payne abundances agree well with the APOGEE measurements for most light, $\alpha$-, and iron-peak elements. The median abundance offsets are generally small, indicating that the abundance scale adopted in this work is well anchored to an independent high-resolution reference.

To investigate whether systematic biases could affect the inferred intrinsic abundance dispersions, Figures~B3 and B4 show the abundance residuals, as functions of effective temperature and metallicity, respectively. Overall, the residuals remain relatively stable across the parameter space occupied by our low-$\alpha$ subgiant sample, with no evidence for large systematic deviations. Magnesium exhibits excellent agreement with APOGEE and shows almost no dependence on either $T_{\rm eff}$ or [Fe/H]. Carbon, oxygen, and silicon display mild systematic trends with both effective temperature and metallicity, but the amplitudes are generally below $\sim0.1$\,dex. Calcium, manganese, and nickel show weak variations over the explored parameter range, while titanium exhibits a larger scatter, particularly at the edges of the sampled parameter space where the number of overlapping stars becomes small. Importantly, none of these trends is sufficiently strong to mimic the much larger ($\gtrsim0.1$\,dex) intrinsic abundance dispersions observed for several neutron-capture elements in the main analysis. These comparisons therefore indicate that the inferred intrinsic abundance dispersions are unlikely to be driven by temperature- or metallicity-dependent systematic effects in the DD-Payne abundance scale.

Taken together, the comparisons with both GALAH DR4 and APOGEE DR19 demonstrate that the DD-Payne abundances used throughout this work are well calibrated against independent high-resolution spectroscopic surveys and do not exhibit significant systematic trends over the relevant stellar parameter range. These tests therefore provide strong support that the intrinsic abundance dispersions reported in this paper primarily reflect astrophysical variations rather than artifacts of the abundance determination procedure.

\begin{figure*}
    \centering
    \includegraphics[width=\textwidth]{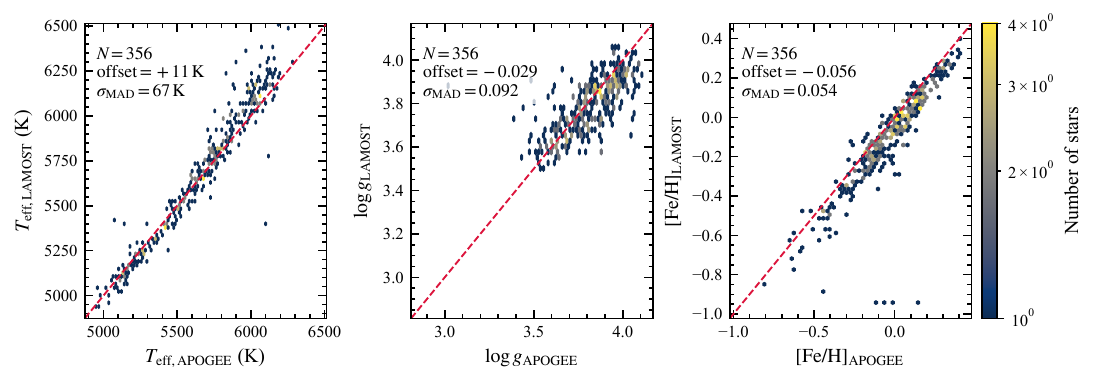}
    \caption{Comparison of stellar parameters between the DD-Payne measurements and APOGEE DR19 for the overlapping subgiant sample. From left to right, the panels show comparisons of $T_{\rm eff}$, $\log g$, and [Fe/H]. The dashed lines indicate one-to-one relations. The median offsets and robust scatters ($\sigma_{\rm MAD}$) are listed in each panel.}
\end{figure*}

\begin{figure*}
    \centering
    \includegraphics[width=\textwidth]{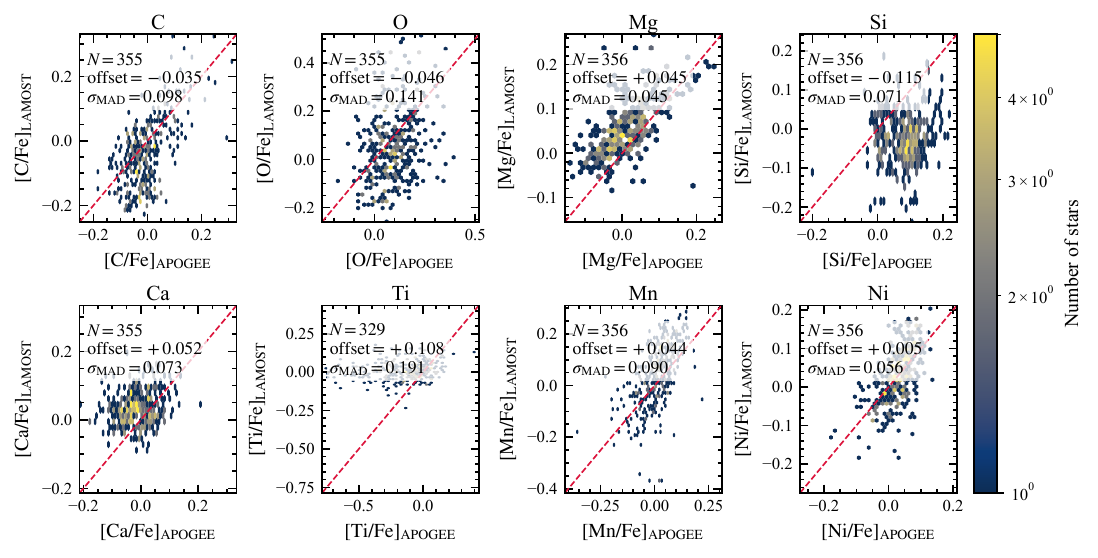}
    \caption{Comparison of elemental abundances between DD-Payne and APOGEE DR19. Each panel shows one abundance ratio measured by the two surveys. The dashed line indicates the one-to-one relation. The number of stars, median abundance offset, and robust scatter ($\sigma_{\rm MAD}$) are reported in each panel.}
\end{figure*}

\begin{figure*}
    \centering
    \includegraphics[width=\textwidth]{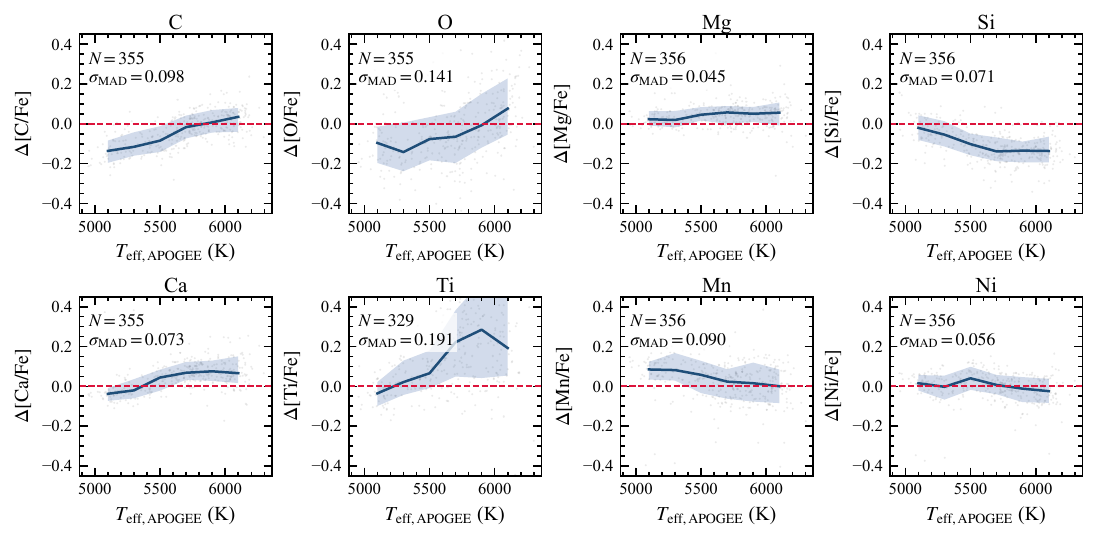}
    \caption{Abundance residuals, defined as $\Delta[X/{\rm Fe}] = [X/{\rm Fe}]{\rm DD\text{-}Payne} - [X/{\rm Fe}]{\rm APOGEE}$, as a function of effective temperature. The solid curves show the running median, while the shaded regions indicate the 16th--84th percentile ranges. Most elements exhibit only weak temperature-dependent systematics across the parameter range of the final sample.}
\end{figure*}

\begin{figure*}
    \centering
    \includegraphics[width=\textwidth]{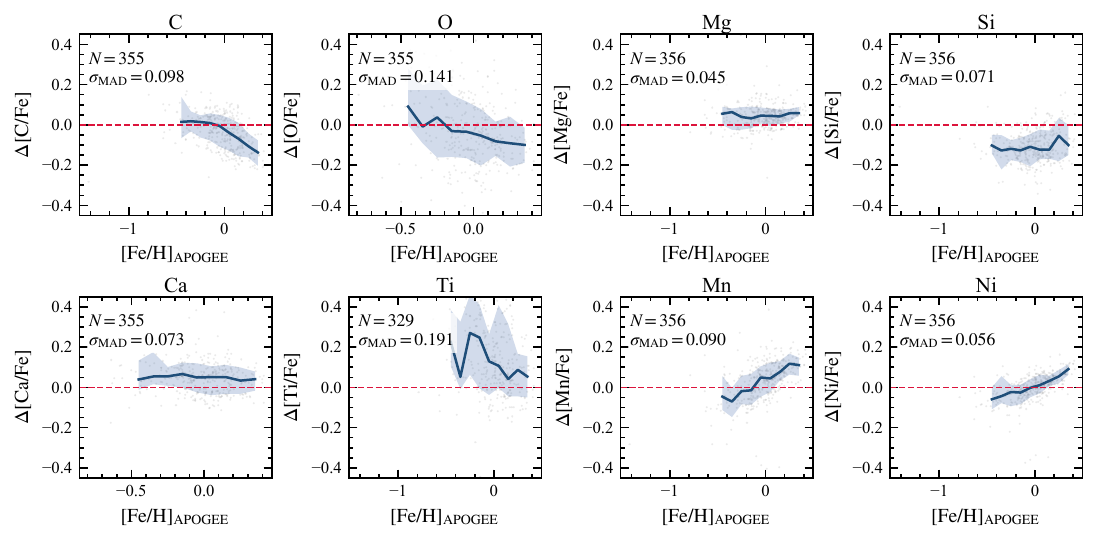}
    \caption{Abundance residuals, defined as $\Delta[X/{\rm Fe}] = [X/{\rm Fe}]{\rm DD\text{-}Payne} - [X/{\rm Fe}]{\rm APOGEE}$, as a function of metallicity. The solid curves show the running median, while the shaded regions indicate the 16th--84th percentile ranges. No strong metallicity-dependent biases are observed for most elements within the metallicity range relevant to this study.}
\end{figure*}

%%%%%%%%%%%%%%%%%%%%%%%%%%%%%%%%%%%%%%%%%%%%%%%%%%

% Don't change these lines
\bsp	% typesetting comment
\label{lastpage}
\end{document}